\documentclass[12pt]{article}
\usepackage{epsfig}
\newcommand{\nn}{\nonumber}
\newcommand{\raw}{\rightarrow}

\newcommand{\be}{\begin{equation}}
\newcommand{\ee}{\end{equation}}
\newcommand{\bea}{\begin{eqnarray}}
\newcommand{\eea}{\end{eqnarray}}

\newcommand{\ev}{ {\rm eV} } 
\newcommand{\evb}{ {\rm eV$^2$} } 
\newcommand{\gev}{ {\rm GeV} }

\newcommand{\PCPV}{
\begin{picture}(22,10)
\put(8,-2){\line(2,1){12}}
\put(0,0){$P_{CP}$}
\end{picture}}
\newcommand{\PCPC}{
\begin{picture}(22,10)
\put(0,0){$P_{CP}$}
\end{picture}}
\newcommand{\PCPVDUE}{
\begin{picture}(24,10)
\put(8,-3){\line(2,1){12}}
\put(0,0){$P_{CP}^{2+2} $}
\end{picture}}
\newcommand{\PCPCDUE}{
\begin{picture}(24,10)
\put(0,0){$P_{CP}^{2+2} $}
\end{picture}}
\newcommand{\PCPVTRE}{
\begin{picture}(24,10)
\put(8,-3){\line(2,1){12}}
\put(0,0){$P_{CP}^{3+1} $}
\end{picture}}
\newcommand{\PCPCTRE}{
\begin{picture}(24,10)
\put(0,0){$P_{CP}^{3+1} $}
\end{picture}}
\begin{document}
%
\thispagestyle{empty}
\begin{flushright}
{hep-ph/0105089}\\
{ROMA1-TH/2001-1313}
\end{flushright}
\vspace*{1cm}
\begin{center}
{\Large{\bf The 2+2 and 3+1 Four-Family Neutrino Mixing at the Neutrino Factory} }\\
\vspace{.5cm}
A. Donini$^{\rm a,}$\footnote{andrea.donini@roma1.infn.it}
and D. Meloni$^{\rm b,}$\footnote{davide.meloni@roma1.infn.it}
 
\vspace*{1cm}
$^{\rm a}$ I.N.F.N., Sezione di Roma I and Dip. Fisica, 
Universit\`a di Roma ``La Sapienza'', P.le A. Moro 2, I-00185, Rome, Italy \\ 
$^{\rm b}$ Dip. Fisica, Universit\`a di Roma ``La Sapienza''
and I.N.F.N., Sezione di Roma I, P.le A. Moro 2, I-00185, Rome, Italy

\end{center}
\vspace{.3cm}
\begin{abstract}
\noindent
We upgrade the study of the physical reach of a Neutrino Factory
in the Four Family Neutrino Mixing scenario taking into account the
latest LSND results that points out how the 3+1 scheme cannot
be completely ruled out within the present experimental data
(although the 2+2 scheme is still the preferred choice when four
neutrinos are considered). A detailed comparison of the physical reach
of the $\nu$-factory in the two schemes is given, with similar
results for the sensitivity to the mixing angles. Huge CP-violating 
effects can be observed in both schemes with a near, $O(10)$ Km, detector of 
$O(10)$ Kton size in the $\nu_\mu \to \nu_\tau$ channel. 
A smaller detector of 1 Kton size can still observe very large effects
in this channel. 
\end{abstract}


\newpage

\section{Introduction}
Indications in favour of neutrino oscillations have been 
obtained both in solar neutrino 
\cite{Cleveland:1998nv,Fukuda:1996sz,Hampel:1999xg,Abdurashitov:1999zd,Suzuki:1999cy}
and atmospheric neutrino 
\cite{Fukuda:1994mc,Becker-Szendy:1995vr,Fukuda:1999ah,Allison:1999ms,Ambrosio:1998wu} 
experiments.
The latest atmospheric neutrino data imply
$\Delta m_{atm}^2 \sim (1.6 - 4) \times 10^{-3}$ \evb \cite{toshito},
whereas the solar neutrino data prefer $\Delta m_{sun}^2 \sim 10^{-10}$
or $10^{-7} - 10^{-4}$ \evb, depending on the particular solution 
for the solar neutrino deficit. 
The LSND data \cite{Athanassopoulos:1998pv,Aguilar:2001ty}, 
on the other hand, would indicate 
a $\bar \nu_\mu \to \bar \nu_e$ oscillation with a third, very distinct, 
neutrino mass difference: $\Delta m_{LSND}^2 \sim 0.3 - 6$ \evb. 
The LSND evidence in favour of neutrino oscillation has not been confirmed 
by other experiments so far. However, the MiniBooNE experiment \cite{Church:1997jc}
will be able to confirm it or not in the near future.
If the LSND results are confirmed we would face
three independent evidence for neutrino oscillations characterized
by squared mass differences quite well separated. 
To explain the whole ensemble of data at least four different light 
neutrino species are needed. The new light neutrino is denoted as sterile 
\cite{Pontecorvo:1968fh}, since it must be an electroweak singlet
to comply with the strong bounds on the $Z^0$ invisible 
decay width \cite{LEPnu}. 
We stress that three massive light neutrinos can not explain all the present
experimental results, as it has been shown with detailed 
calculations in \cite{Fogli:1999zq}.

There are two, very different, classes of four neutrino spectra: 
three almost degenerate neutrinos and an isolated fourth one, 
or two pairs of almost degenerate neutrinos divided by the large
LSND mass gap. The two classes of mass spectra are usually called 
the 3+1 and 2+2 schemes \cite{Caldwell:1993kn}
, respectively. All the present experimental 
evidence for neutrino oscillations have been combined in the literature
in order to identify which of the two classes of mass spectra
better adapts to the data. The experimental results were strongly
in favour of the 2+2 scheme \cite{Bilenkii:1999ny} until the latest
LSND results have been presented in June 2000, \cite{mills}
(see \cite{Aguilar:2001ty}). 
The new analysis of the experimental data results in a shift of the
allowed region towards smaller values of the mixing angle, 
$\sin^2 (2 \theta_{LSND})$, reconciling the 3+1 scheme with exclusion 
bounds coming from CDHS \cite{Dydak:1984zq}, CCFR \cite{Stockdale:1985ce} 
and Bugey \cite{Declais:1995su}.
Although the 2+2 scheme is still favoured\footnote{A novel 
Bayesian analysis
of the exclusion bounds, in the spirit of \cite{Bilenkii:1999ny}
has been presented in \cite{Grimus:2001mn}, claiming that the 3+1
scheme is allowed at the 99 \% CL only, but not at the 95 \% CL.}, 
the 3+1 scheme is at present marginally compatible with the data, 
\cite{Barger:2000ch,Giunti:2001ur,Peres:2001ic,Yasuda:2000dc}. 
However, the 2+2 and the 3+1 scheme face the upcoming experiments
on totally different footing: if MiniBooNE disconfirms LSND, the 2+2
scheme is falsified. On the contrary, it is not possible to falsify
the 3+1 scheme: we can always consider an extension of the Standard Model 
with three light neutrinos and a fourth sterile 
one, separated by some squared mass difference, $\Delta m^2_{(1,2,3)-4}$. 
The implication of a negative result of MiniBooNE is just
$\Delta m^2_{(1,2,3)-4} \neq \Delta m^2_{LSND}$.

The specific form of the neutrino mass spectrum appears, therefore, one 
of the (many) open questions related to the lepton sector of the
Standard Model.

Four neutrino oscillations imply a Maki-Nakagawa-Sakata (MNS) $4 \times 4$
mixing matrix, with six rotation angles $\theta_{ij}$ and three phases $\delta_i$
(for Majorana neutrinos, three additional phases are allowed, but
they are not testable in oscillation experiments, and therefore will not
be considered here). A Neutrino Factory \cite{Geer:1998iz,DeRujula:1999hd}
seems to be the best option to explore this huge parameter space. 
The $\mu^\pm$-decay into the straight
section of a muon storage ring should produce a very intense and pure
neutrino beam. The rich flavour content (50 \% of 
$\nu_\mu (\bar \nu_\mu)$ and 50 \% of $\bar \nu_e (\nu_e)$ are 
simultaneously produced), finally, makes the Neutrino Factory well suited 
for precision studies of the MNS mixing matrix, hopefully including 
the discovery of leptonic CP-violation \cite{Albright:2000xi,Blondel:2000gj}.
The following scheme reminds that at the Neutrino Factory $\mu$ and $\tau$
appearance channels can also be used, in combination with the $\mu$ and $e$
disappearance experiments. 
\bea
                       & e^-, \tau^-     &          \nn \\
                       & \uparrow        &          \nn \\
                       & \nu_e,\nu_\tau  &          \nn \\      
                       & \uparrow        &          \nn \\
\mu^-  \rightarrow e^- & \nu_\mu         & \bar \nu_e 
\qquad \rightarrow \qquad \mu^-,e^+ \\
                       &                 &\downarrow  \nn \\
                       &                 & \bar \nu_\mu, \bar \nu_\tau \nn \\
                       &                 & \downarrow \nn \\
                       &                 & \mu^+, \tau^+ \nn
\eea
In \cite{Cervera:2000kp,Donini:2000ky,Burguet-Castell:2001ez} 
the ``wrong-sign muon'' channel ($\mu^+$ appearance in a $\mu^-$ beam) 
has been shown to be extremely useful to explore the parameter space 
of three-family neutrino mixing, 
with particular interest in the measure of the (single) CP-violating phase, 
thus deserving the nickname of ``golden measurement'' at a Neutrino Factory. 
This has to be compared with a conventional beam experiment (using 
muon neutrinos from pion decay), such as K2K or the approved FermiLab 
to Soudan long baseline experiment. In these experiments, mainly the 
$\mu$-disappearance channel is exploited. 

In \cite{Donini:1999jc,Donini:2000he} it was shown that a Neutrino Factory 
with $10 - 50$ GeV muons can attain $\sin^2 (\theta_{ij})$ as low as 
$10^{-5} - 10^{-3}$ for $\Delta m^2_{LSND} \in [10^{-1}, 10^1]$ \evb. Moreover, it was found
that sizeable CP-violating effects can be observed in the 
$\nu_\mu \to \nu_\tau$ channel with a 1 Kton detector located at $O (10)$ Km.
This analysis has been performed in the 2+2 scheme, the only allowed
at that moment. The first motivation for this paper is, therefore,
the comparison of what has been found for the 2+2 scheme with 
the same kind of analysis in the, by now marginally allowed, 3+1 scheme.
A better understanding of the oscillation probability structure 
is a natural by-product of this analysis, both for the CP-conserving
and the CP-violating part. In particular, we found a simple argument
that shows how the $\nu_\mu \to \nu_\tau$ channel is the best suited one
for CP-violation experiments in four neutrino mixing, to be 
compared with the three-family mixing where $\nu_e \to \nu_\mu$ happens 
to be optimal. Finally, the same argument justifies the 
loss in sensitivity to small mixing angles in the 3+1 scheme with 
respect to the 2+2 scheme. 

The considered set-up is, as in \cite{Donini:1999jc,Donini:2000he}, 
a Neutrino Factory with $2 \times 10^{20} \mu^+$ and $\mu^-$ decaying
in the straight section of a $10 - 50$ GeV muon storage ring per year,
and five years of data taking. Muon 
energies in this range are at present under discussion. The higher 
energy range allows a good background rejection \cite{Cervera:2000vy};
moreover, the integrated flux times the cross-section increase 
with $E_\mu$. A high-energy Neutrino Factory seems therefore the 
best option, with the energy mainly limited by cost considerations.
However, although the total number of charged leptons
into the detector increases with the parent muon energy, 
the flux of low energy neutrinos decreases. If low energy neutrinos 
are needed (for example, to study CP-violating observables
strongly reducing the matter effects \cite{Minakata:2000ee,Koike:2000jf}),
this reduction in the flux should be taken into account.
 
The mixing angles that relate neutrino mass eigenstates with an LSND mass difference
can be studied in short baseline experiments, $L \sim 1$ Km. A small detector 
with $O (1)$ ton mass is, therefore, well suited to study the whole
gap-crossing parameter space (due to the large neutrino flux that 
illuminates the detector). To take full advantage of the rich flavour
content of the beam, this detector should be equipped with $\tau$-tracking
and ($\mu , \tau$) charge identification. If CP-violating observables
are considered, the best option is a larger, $O (10)$ Kton detector
located at $O (10)$ Km down the source. This set-up is equally 
powerful both for the 2+2 and the 3+1 schemes, and should be 
compared with the typical set-up needed when three-family neutrino
mixing is considered. 

We also try to answer to the following question: is it possible to explore the 
whole parameter space with a different detector, with no $\tau$-tracking, 
but taking full advantage of the energy dependence of the transition probabilities, 
in the spirit of \cite{Cervera:2000kp}? We focus on the $\nu_e \to \nu_\mu$
channel in the 3+1 scheme, with a realistic 10 Kton magnetized iron detector 
of the type presented in \cite{Cervera:2000vy} located at $L = 40$ Km down the source. 
The Neutrino Factory is run with $2 \times 10^{20}$ useful muons per year for 5 
operational years for both muon polarities, at $E_\mu = 50$ \gev, with a detector energy
resolution of $\Delta E_\nu = 10$ \gev. A detailed estimate of the backgrounds and 
detection efficiencies of the considered detector has been presented in 
\cite{Cervera:2000kp}. The energy dependence of the oscillation probabilities 
could in principle help in the measurement of two gap-crossing 
angles, or one angle and a CP-violating phase, at a time. In the latter case, 
we find results similar to those in \cite{Cervera:2000kp}: we can easily
reconstruct the phase and the angle at the same time, with an error of tens
of degrees on $\delta_i$ and of tenths of degree on the angle. 
However, it seems extremely difficult to measure two gap-crossing angles at a time
in the $\nu_e \to \nu_\mu$ channel. Our conclusion is that to fully explore
the parameter space of the four-family model a detector with $\tau$-tracking is needed. 

The paper is organized as follows: in Sect. \ref{param} we introduce
our parametrization of the MNS mixing matrix both for the 2+2 and 3+1
schemes; in Sect. \ref{bounds} the present bounds on the mixing angles coming 
from existing experiments are given; in Sect. \ref{setup} we describe the Neutrino Factory
and detector setup; in Sect. \ref{sensi} we present
our results for the sensitivity of the Neutrino Factory to the 
mixing angles (in the case of no CP violation), comparing the
2+2 and 3+1 schemes; in Sect. \ref{cp-viola} we extend our analysis
to the CP-violating observables; in Sect. \ref{fit4in4} we 
study the possibility of measuring two gap-crossing angles 
or one angle and a phase at a time exploiting the energy dependence
of the transition probabilities; in Sect. \ref{concl} we eventually 
draw our conclusions. 

\section{The Four Neutrino mixing matrix}
\label{param}

\begin{figure}[h!]
\begin{center}
\mbox{\epsfig{file=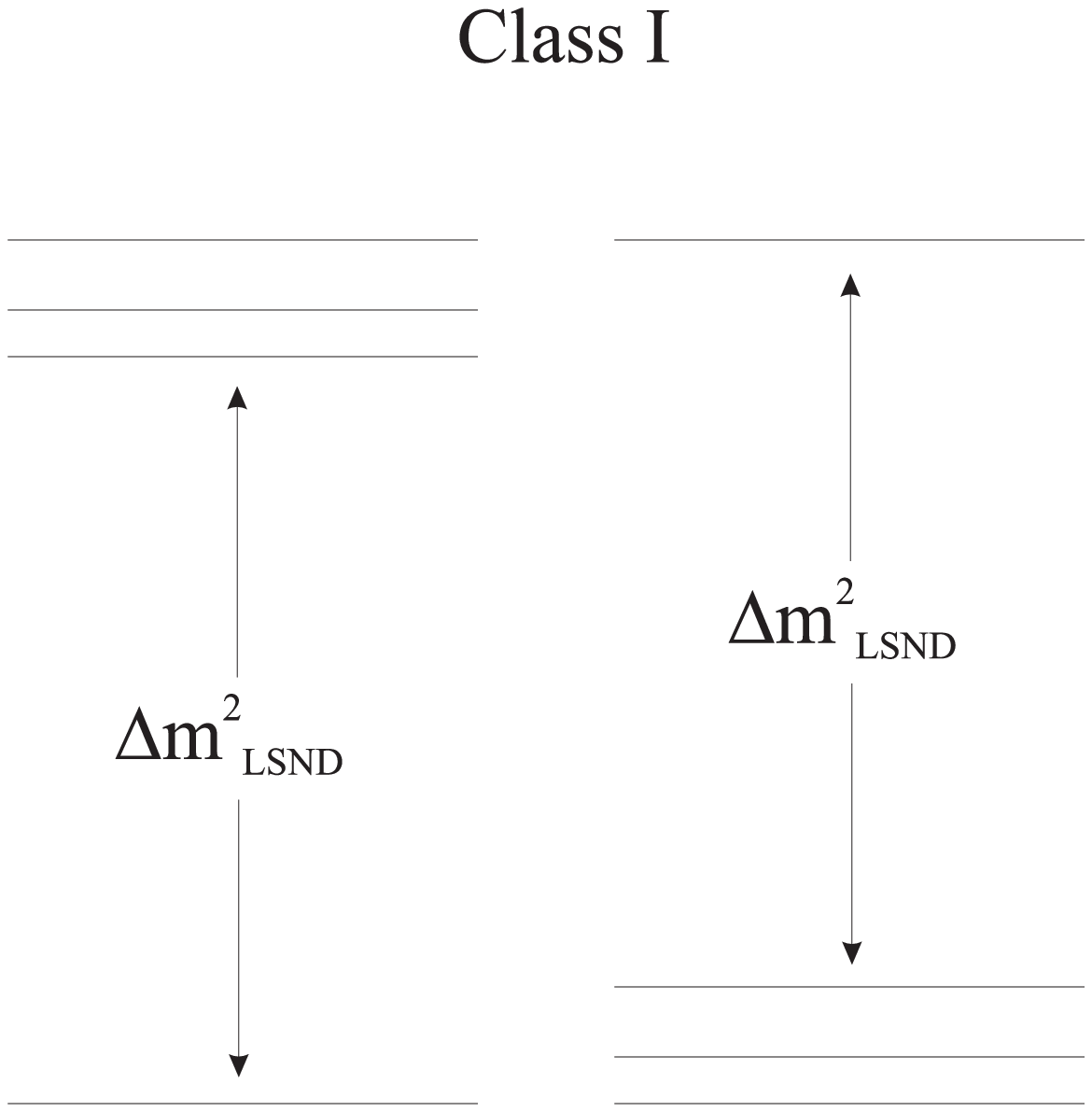,height=5.cm}} \hspace{0.7truecm}
\mbox{\epsfig{file=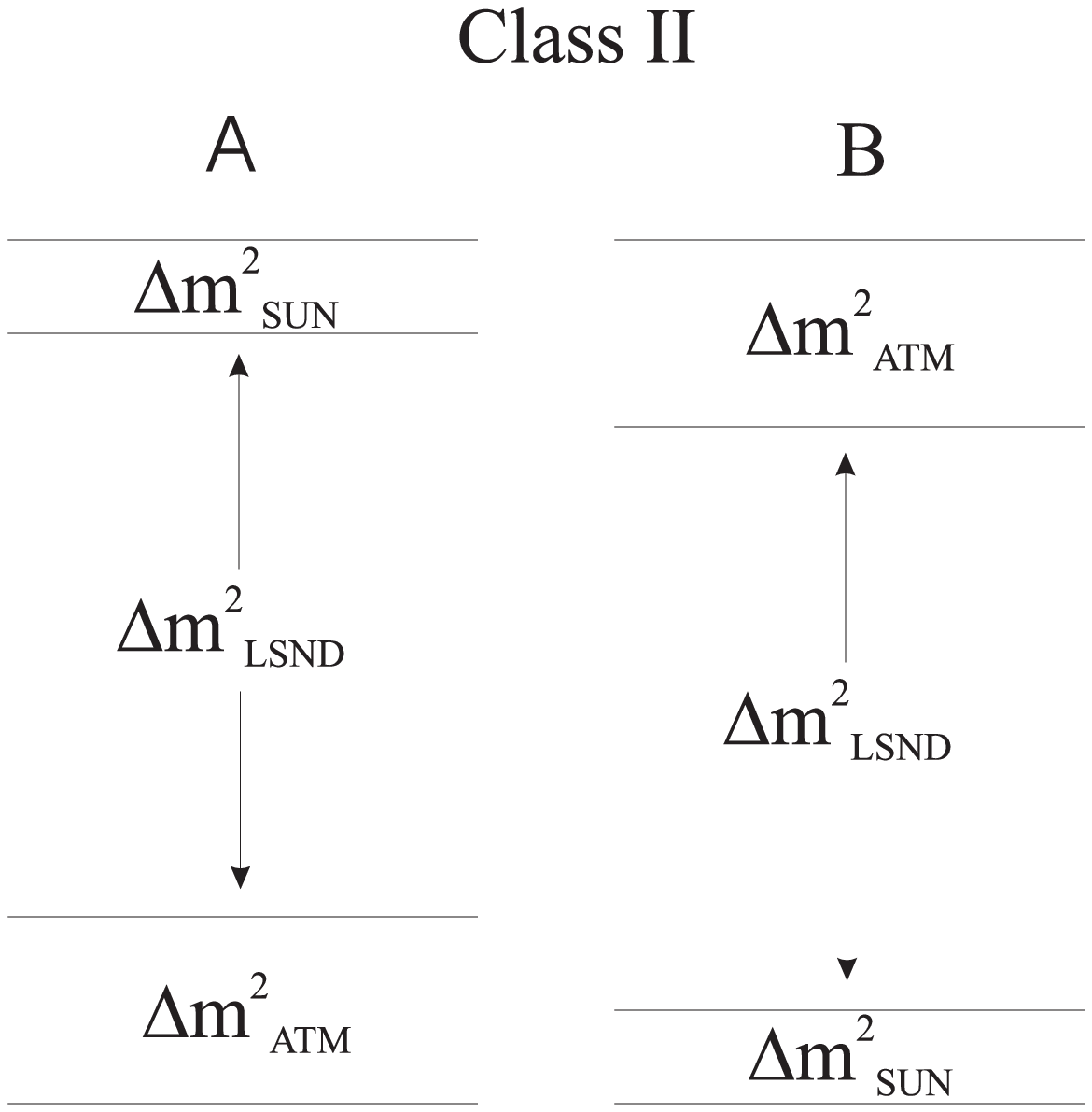,height=5.cm}}
\end{center}
\vspace*{0.5cm}
\caption{{\it Different types of four family neutrino mass spectrum: 3+1
scenarios (left); 2+2 scenarios (right).}}
\label{fig:classes}
\end{figure}

When four neutrinos are considered, two very different classes of 
mass spectrum are possible: three almost degenerate (mainly active) 
neutrinos, accounting for the solar and atmospheric oscillations, 
separated from the fourth (mainly sterile) one by the large
LSND mass difference, $\Delta m^2_{LSND}$; or, 
two almost degenerate neutrino pairs, accounting respectively for the
solar and atmospheric oscillations, separated by the LSND mass gap.
The two mass spectrum classes are depicted in Fig. \ref{fig:classes}.
We refer to these possibilities as 3+1 and 2+2 scenarios. 
There are four 3+1 and two 2+2 scenarios depending on the 
specific ordering of the mass differences. Notice that the
intriguing hierarchical and inverted hierarchical mass spectrum 
are 3+1 scenarios. 

It has been shown in \cite{Bilenkii:1999ny} that the combined analysis
of solar, atmospheric and LSND data disfavours the 3+1 scheme. 
For this reason, the 2+2 scheme has been carefully studied in 
the recent literature (see, for example, \cite{Giunti:2000wt,Yasuda:2000de} 
and references therein). Consider for definiteness the rightmost
scenario, namely the lower pair accounting for the solar neutrino deficit
and the higher pair for the atmospheric one (the other possibility
directly follows by changing the sign of the $\Delta m^2_{LSND}$). 
The $\nu_\mu$ is therefore
in the heavier pair and the $\nu_e$ in the lighter one. 
Is the sterile neutrino, $\nu_s$, responsible for the observed atmospheric
oscillations or for the solar neutrino deficit? 
The latest SuperKamiokande results for the atmospheric neutrinos\footnote{ 
In particular, the zenith angle distribution of the upward-going muons 
\cite{Lipari:1998rf}, the partially contained multi-ring events and 
the neutral current data sample.} disfavours total conversion of $\nu_\mu$ 
into the sterile neutrino at 99 \% CL \cite{kajita}. 
Moreover, the conversion of $\nu_e$ solar neutrinos into active neutrinos
gives better global fits of the experimental data, with respect
to active-into-sterile conversion \cite{suzuki}.
Although partial conversion of $\nu_\mu \to \nu_s$ or $\nu_e \to \nu_s$
is not excluded (with a sterile component in the atmospheric oscillation
as large as 50 \%, \cite{Fogli:2001ir}), the present solar and atmospheric
data suggest active-to-active oscillations. The active-to-sterile
oscillation should therefore be responsible only for the LSND
results. This scenario appears quite unnatural in the framework of 
the 2+2 scheme.

The latest analysis of the LSND data \cite{mills}, however, shows
a shift of the allowed region for the LSND two-family-equivalent
mixing angle, $\sin^2 (2 \theta_{LSND})$, towards smaller values. 
This reconciles the 3+1 scheme with the exclusion 
bounds coming from CDHS \cite{Dydak:1984zq}, CCFR \cite{Stockdale:1985ce} 
and Bugey \cite{Declais:1995su}. In the 3+1 scheme, the three
almost degenerate neutrinos are mainly active and the separated
fourth is mainly sterile; the gap-crossing mixing angles are generally
small. In this scenario, the interpretation of solar and atmospheric
oscillations as active-to-active and LSND as active-to-sterile
naturally arises: this scheme is a deformation of the Standard
Model, slowly decoupling as the gap-crossing mixing angles become
smaller and smaller (and thus, the sterile becomes irrelevant). 

Given $n$ light neutrino species, the most general mixing matrix
is an $n \times n$ unitary matrix \cite{Maki:1962mu}, $U_{MNS}$. 
For $n = 4$, the MNS matrix contains six independent rotation angles $\theta_{ij}$ 
and three (if neutrinos are Dirac fermions) or six (if neutrinos are Majorana
fermions) phases $\delta_i$. However, oscillation experiments are only
sensitive to the first three phases, the effect of the
Majorana phases being suppressed by factors of $m_\nu / E_\nu$. 
The Majorana or Dirac nature of neutrinos
can thus be tested only in $\Delta L = 2$ transitions such as
neutrino-less double $\beta$-decay \cite{Bilenky:2001xq}.
In the following analysis, with no loss in generality,
we will restrict ourselves to Dirac-type neutrinos only. 
We consider a hierarchical 3+1 spectrum and a class II-B 2+2 spectrum, 
for definiteness.

This large parameter space (6 angles and 3 phases, to be compared
with the standard three-family mixing case of 3 angles and 1 phase)
is actually reduced to a smaller subspace whenever some of the 
mass differences become negligible. Consider the measured hierarchy in the 
mass differences, 
\be
\Delta m^2_{sol} \ll \Delta m^2_{atm} \ll \Delta m^2_{LSND} \, , 
\ee
and define
\be
\Delta_{ij} = \frac{\Delta m^2_{ij} L}{4 E_\nu} \, .
\label{def:deltaij}
\ee
At short distance, $L = O(1)$ Km, for neutrinos up to $O (10)\, \gev$, 
\bea
\Delta_{sol} \; , \; \Delta_{atm} \ll 1 \, , \nn \\
\Delta_{LSND} = O(1) \, .
\eea
Therefore, it is natural at short distances to neglect the solar
and atmospheric mass difference and to work in a reduced parameter
space. This approximation is called ``one-mass dominance'' \cite{DeRujula:1980yy}.
In the 2+2 scheme, neglecting the smaller mass differences
implies that rotations in the $(1-2)$ and $(3-4)$ planes are irrelevant.
Thus, it is not possible to measure the rotation angles in these planes
in oscillation experiments. Two CP-violating phases also become irrelevant, 
and therefore the reduced parameter space in the 2+2 scheme contains
4 rotation angles and 1 phase. 
In the 3+1 scheme, neglecting the solar and atmospheric mass differences
implies that rotations in the whole three-dimensional subspace $(1-2-3)$
are irrelevant for oscillation experiments, and the physical parameter
space contains just three rotation angles and no phases. 
When considering CP-violating phenomena, however, at least two mass
differences should be taken into account: in this case
we neglect the solar mass difference and consider the atmospheric
mass difference a perturbation. This is called ``two-mass dominance''
approximation. In this approximation, regardless of the scheme, 
the parameter space contains 5 angles and 2 phases. The number of independent
parameters of the MNS mixing matrix in four neutrino models is summarized in 
Tab. \ref{tab:parameters}.
\begin{table} [h]
{\small
\center 
\begin{tabular}{||c|c|c|c||}
\hline\hline 
         & Angles & Dirac CP-phases & Majorana CP-phases \\
\hline\hline
& & & \\
Majorana $\nu$'s &   6   &   3   &  3   \\
& & & \\
\hline 
& & & \\
Dirac $\nu$'s &   6   &   3   &  0   \\
& & & \\
\hline
& & & \\
Dirac $\nu$'s   &   5   &   2   &  0   \\
$\Delta m_{12}^2 = 0$ & & & \\
& & & \\
\hline
& & & \\
2+2  &   &    &  \\
Dirac $\nu$'s   &   4   &   1   &  0   \\
$\Delta m_{12}^2 = \Delta m_{34}^2 = 0$ & & & \\
& & & \\
\hline
& & & \\
3+1  &   &    &  \\
Dirac $\nu$'s &   3   &   0   &  0  \\
$\Delta m_{12}^2 = \Delta m_{34}^2 = 0$ & & & \\
& & & \\
\hline \hline
\end{tabular}
\caption{ {\it Parameter space in four neutrino models: for Dirac neutrinos we consider
the general case (three non-zero mass differences) and the one- and two-mass 
dominance approximations; for Majorana neutrinos we consider the general case only.}}
\label{tab:parameters}
}
\end{table}

A generic rotation in a four dimensional space can be obtained by performing six
different rotations $U_{ij}$ in the $(i-j)$ plane, resulting in plenty of different
parametrizations of the mixing matrix (and still not taking into account the three 
CP-violating phases). However, in \cite{Donini:1999jc,Donini:2000he} was shown how the 
one-mass dominance and two-mass dominance approximations can be implemented in 
a trasparent way (in the sense that only the physical parameters appear in the 
CP-conserving and CP-violating oscillation probabilities). 
A convenient parametrization of the mixing matrix is that in which the 
rotation matrices corresponding to the most degenerate pairs of eigenstates are 
located at the extreme right. 
If the eigenstates $i$ and $j$ are degenerate and the matrix $U_{ij}$ 
is the rightmost one, the corresponding angle $\theta_{ij}$ automatically disappears 
from the oscillation probabilities, and the parameter space gets reduced to the truly 
observable angles and phases. If a different ordering of the 
rotation matrices is taken, no angle disappears from the oscillation formulae, 
and a parameter redefinition would be necessary to reduce the parameter space 
to the observable sector. 

In the 2+2 scheme, the following parametrization was adopted in \cite{Donini:1999jc}
implementing the previous argument:
\be
U_{MNS} = U_{14} (\theta_{14}) \; U_{13} (\theta_{13}) \; U_{24} (\theta_{24}) \; 
    U_{23} (\theta_{23}\, , \, \delta_3) \; U_{34} (\theta_{34}\, , \, \delta_2) \; 
    U_{12} (\theta_{12}\, , \, \delta_1).
\label{2+2param}
\ee 
In the one-mass dominance approximation, the unphysical angles and phases 
$(\theta_{12}, \delta_1)$ and $(\theta_{34}, \delta_2)$ automatically decouple. 
The oscillation probabilities in the appearance channels are\footnote{
In what follows, we separate the CP-even terms from the CP-odd ones: \par
$P (\nu_\alpha \to \nu_\beta) = \PCPC (\nu_\alpha \to \nu_\beta)
+ \PCPV (\nu_\alpha \to \nu_\beta)$.}: 
\bea
\PCPCDUE(\nu_e \to \nu_\mu) &=& 
4 c^2_{13} c^2_{24} c^2_{23} s^2_{23} \,
\sin^2 \left ( \frac{\Delta m_{23}^2 L}{4 E} \right ) \ , 
\label{eq:emu22} \\
\PCPCDUE(\nu_e \to \nu_\tau) & = & 
4 c^2_{23} c^2_{24} \left [ (s^2_{13} s^2_{14} s^2_{23} + c^2_{14} c^2_{23} s^2_{24})
\right . \nn \\
&-& \left . 2 c_{14} s_{14} c_{23} s_{23} s_{13} s_{24} \cos \delta_3
\right ] \, \sin^2 \left ( \frac{\Delta m_{23}^2 L}{4 E} \right )  \ , 
\label{eq:etau22} \\
\PCPCDUE(\nu_\mu \to \nu_\tau) & = & 
4 c^2_{23} c^2_{13} \left [ (s^2_{13} s^2_{14} c^2_{23} + c^2_{14} s^2_{23} s^2_{24})
\right . \nn \\
&+& \left . 2 c_{14} s_{14} c_{23} s_{23} s_{13} s_{24} \cos \delta_3
\right ] \, \sin^2 \left ( \frac{\Delta m_{23}^2 L}{4 E} \right ) \ ;
\label{eq:mutau22}
\eea
in the disappearance channels are:
\bea
\PCPCDUE(\nu_\mu \to \nu_\mu) & = & 1
 - 4 c^2_{13} c^2_{23} ( s^2_{23} + s^2_{13} c^2_{23} ) \,
       \sin^2 \left ( \frac{\Delta m_{23}^2 L}{4 E} \right ) \ ,
\label{eq:mumu22} \\
\PCPCDUE(\nu_e \to \nu_e) & = & 1 - 4 c^2_{23} c^2_{24} 
( s^2_{24} + s^2_{23} c^2_{24} ) \,
\sin^2 \left ( \frac{\Delta m_{23}^2 L}{4 E} \right ) \ .
\label{eq:ee22}
\eea
Notice that the physical phase $\delta_3$ appears in the CP-conserving
transition probabilities in a pure cosine dependence. No CP-odd observable
can be built out of the oscillation probabilities in this approximation
in spite of the existence of a physical phase in the mixing matrix. 

In the two-mass dominance approximation, new CP-violating terms arise. Expanding
the probabilities at first order in $\Delta_{atm}$, we get\footnote{In 
\cite{Donini:1999jc,Donini:2000he} some misprints were present in the 
$\nu_e \to \nu_\tau$ formula that have been corrected here.}:
 \bea
\PCPVDUE(\nu_e \to \nu_\mu) & = & 
    - 8 c^2_{13} c^2_{23} \, c_{24} c_{34} s_{23} s_{24} s_{34} 
     \sin (\delta_2 + \delta_3) \nn \\ 
& \times &
     \left( {{ \Delta m^2_{34} L }\over{4 E_\nu} } \right) \,
     \sin^2 \left (\frac{\Delta m_{23}^2 L}{4 E_\nu} \right) \ , 
\label{eq:emu22cp} \\
\PCPVDUE(\nu_e \to \nu_\tau) & = & 8 c_{23} c_{24} 
      \left \{ 
      c_{23} c_{34} s_{23} s_{24} s_{34} 
      ( c^2_{14} - s^2_{13} s^2_{14} ) \ \sin (\delta_2 + \delta_3) \right . \nn \\
& + & 
      c_{14} c_{34} s_{13} s_{14} s_{34} \left [ 
      (s^2_{24} - s^2_{23}) \ \sin \delta_2 - 
      s^2_{23} s^2_{24} \ \sin(\delta_2 + 2\delta_3) \right ] \nn \\
& + & \left .
      c_{14} c_{24} s_{13} s_{14} s_{23} s_{24} 
      (c^2_{34} - s^2_{34}) \ \sin \delta_3  \right \} \nn \\
& \times &
      \left( {{ \Delta m^2_{34} L }\over{4 E_\nu} } \right ) \, 
      \sin^2 \left( \frac{\Delta m_{23}^2 L}{4 E_\nu} \right ) \ , 
\label{eq:etau22cp} \\
\PCPVDUE(\nu_\mu \to \nu_\tau) & = & 
      - 8 c^2_{13}  c^2_{23} \, c_{14} c_{24} c_{34} s_{34} \left [ 
      c_{23} s_{13} s_{14} \ \sin \delta_2 \ + 
      c_{14} s_{23} s_{24} \ \sin (\delta_2 + \delta_3) \right ] \nn \\ 
& \times & \left( {{ \Delta m^2_{34} L }\over{4 E} } \right ) \,
      \sin^2 \left( \frac{\Delta m_{23}^2 L}{4 E} \right ) \ .
\label{eq:mutau22cp}
\eea
Two distinct phases, $\delta_2$ and $\delta_3$, appear in these formulae 
in a characteristic sine dependence which is the trademark of CP-violating
observables. CP-violating effects can only be measured in appearance channels, 
whereas the disappearance channels $\nu_e \to \nu_e$ and $\nu_\mu \to \nu_\mu$
are only sensitive to the CP-even parameters, 
\be
\PCPV(\nu_e \to \nu_e)  = \PCPV(\nu_\mu \to \nu_\mu)  = 
                          \PCPV(\nu_\tau \to \nu_\tau) = 0 \ .
\ee

In the 3+1 scheme, the following parametrization shares the same virtues of 
eq.~(\ref{2+2param}):
\be
U_{MNS} = U_{14} (\theta_{14}) \; U_{24} (\theta_{24}) \; U_{34} (\theta_{34}) \; 
    U_{23} (\theta_{23}\, , \, \delta_3) \; U_{13} (\theta_{13}\, , \, \delta_2) \; 
    U_{12} (\theta_{12}\, , \, \delta_1).
\label{3+1param}
\ee 
This parametrization has the additional advantage that the three-family model
mixing matrix in its standard form can be immediately 
recovered when $\theta_{i4} = 0$. 
For small gap-crossing angles $\theta_{i4}$, we expect slight modification
with respect to the three-family model.

In the one-mass dominance approximation, the unphysical angles and phases 
$(\theta_{12}, \delta_1)$, $(\theta_{13}, \delta_2)$ and $(\theta_{23}, \delta_3)$ 
automatically decouple. The oscillation probabilities in the appearance channels are: 
\bea
\PCPCTRE(\nu_e \to \nu_\mu) &=& 4 c^2_{24} c^4_{34} s^2_{14} s^2_{24} \,
\sin^2 \left ( \frac{\Delta m_{34}^2 L}{4 E} \right ) \ , 
\label{eq:emu31} \\
\PCPCTRE(\nu_e \to \nu_\tau) & = & 4 c^2_{24} c^2_{34} s^2_{14} s^2_{34} \,
\sin^2 \left ( \frac{\Delta m_{34}^2 L}{4 E} \right )  \ , 
\label{eq:etau31} \\
\PCPCTRE(\nu_\mu \to \nu_\tau) & = & 4 c^2_{34} s^2_{24} s^2_{34} \,
\sin^2 \left ( \frac{\Delta m_{34}^2 L}{4 E} \right ) \ ;
\label{eq:mutau31} 
\eea
in the disappearance channels are:
\bea
\PCPCTRE(\nu_\mu \to \nu_\mu) & = & 1
 - 4 c^2_{34} s^2_{24} ( c^2_{24} + s^2_{24} s^2_{34} ) \,
       \sin^2 \left ( \frac{\Delta m_{34}^2 L}{4 E} \right )  \ ,
\label{eq:mumu31} \\
\PCPCTRE(\nu_e \to \nu_e) & = & 1 - 4 c^2_{24} c^2_{34} s^2_{14} (1-s^2_{14}
c^2_{24} c^2_{34}) \,
\sin^2 \left ( \frac{\Delta m_{34}^2 L}{4 E} \right ) \, .
\label{eq:ee31} 
\eea
As already stressed, angles and phases in the three-dimensional physically
irrelevant subspace are not present in these formulae. 

Finally, in the two-mass dominance approximation we get, expanding at first
order in $\Delta_{atm}$: 
\bea
\PCPVTRE(\nu_e \to \nu_\mu) & = & 8~c^2_{34} c_{23} c_{24} s_{14} s_{24} \, 
     \left \{ - c_{13} s_{14} s_{23} s_{34} \sin \delta_2 \right . \nn \\ 
&+&  c_{14} s_{13} \left . \left [
     c_{13} c_{23} s_{24} s_{34} \sin \delta_3 + c_{24} s_{23} 
     \sin (\delta_2 - \delta_3) \right ] \right \} \nn \\
& \times &
     \left( {{ \Delta m^2_{23} L }\over{4 E_\nu} } \right) \,
     \sin^2 \left (\frac{\Delta m_{34}^2 L}{4 E_\nu} \right) \ , 
\label{eq:emu31cp} \\
\PCPVTRE(\nu_e \to \nu_\tau) & = & 8~c^2_{34} c_{13} c_{23} c_{24} s_{14} s_{34} 
      \left [ s_{14} s_{23} s_{24} \sin \delta_2 - c_{14} c_{23} s_{13} \sin \delta_3
   \right ] \nn \\
& \times &
      \left( {{ \Delta m^2_{23} L }\over{4 E_\nu} } \right) \,
      \sin^2 \left (\frac{\Delta m_{34}^2 L}{4 E_\nu} \right) \ , 
\label{eq:etau31cp} \\
\PCPVTRE(\nu_\mu \to \nu_\tau) & = & 
     - 8~c^2_{34} c_{23} c_{24} s_{23} s_{24} s_{34} \sin \delta_2 \nn \\
& \times &
     \left( {{ \Delta m^2_{23} L }\over{4 E} } \right ) \, 
      \sin^2 \left( \frac{\Delta m_{34}^2 L}{4 E} \right ) .
\label{eq:mutau31cp} 
\eea

These formulae, both in the 2+2 and the 3+1 scheme, will be used in
Sect. \ref{sensi} and Sect. \ref{cp-viola} 
to explore the parameter space of the four-family model
at the Neutrino Factory. 

\subsection{Experimental bounds on the gap-crossing angles}
\label{bounds}

We recall here the bounds on the rotation angles
and mass differences coming from the existing experiments. The Bugey and CHOOZ 
experiments \cite{Declais:1995su,Apollonio:1999ae}
give strong upper limit to the $\nu_e \to \nu_e$ disappearance
two-family equivalent mixing angle. In two families, 
\be
\
\PCPC (\nu_e \to \nu_e) = 1 - \sin^2 (2 \theta)_{exp} \, 
\sin^2 \left ( \frac{\Delta m^2_{LSND} L}{4 E} \right )
\ee
with $\sin^2 (2 \theta)_{exp} \leq 0.2$ in the LSND-allowed region. 
The positive result from LSND gives a lower limit on the 
$\nu_e \to \nu_\mu$ two-family--equivalent mixing angle, 
\be
\PCPC (\nu_\mu \to \nu_e) = \sin^2 (2 \theta)_{LSND} \, 
\sin^2 \left ( \frac{\Delta m^2_{LSND} L}{4 E} \right )
\ee
with $10^{-3} \leq \sin^2 (2 \theta)_{LSND} \leq 1$.
These bounds, jointly with the negative results from Karmen2 \cite{Kleinfeller:2000em} 
and previous experiments such as CDHS and CCFR \cite{Dydak:1984zq,Stockdale:1985ce}, 
must be interpreted in the 2+2 and 3+1 scheme, 
extracting informations on the gap-crossing angles and mass differences.

\begin{figure}[h!]
\begin{center}
\mbox{\epsfig{file=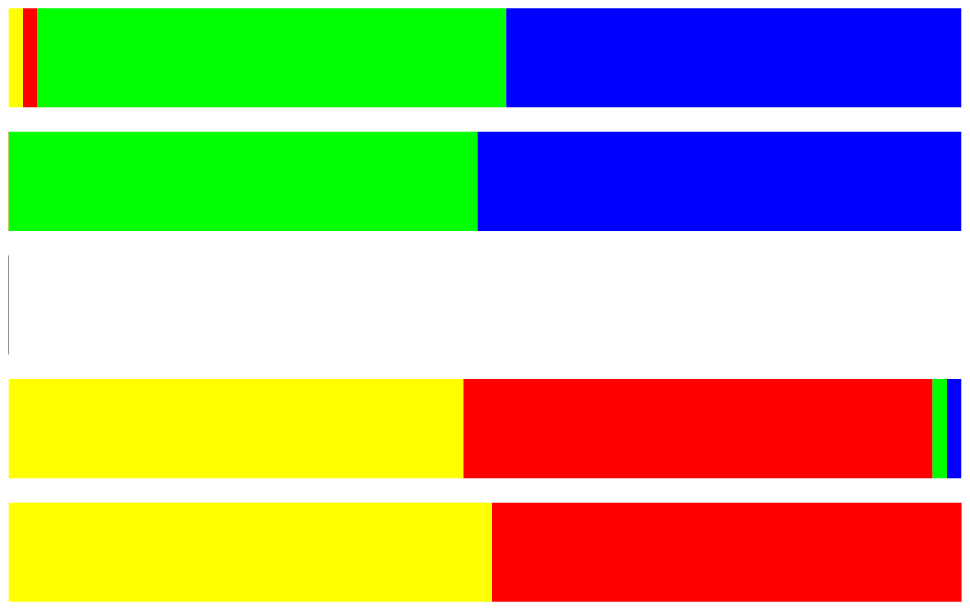,width=5.cm}} \hspace{0.7truecm}
\mbox{\epsfig{file=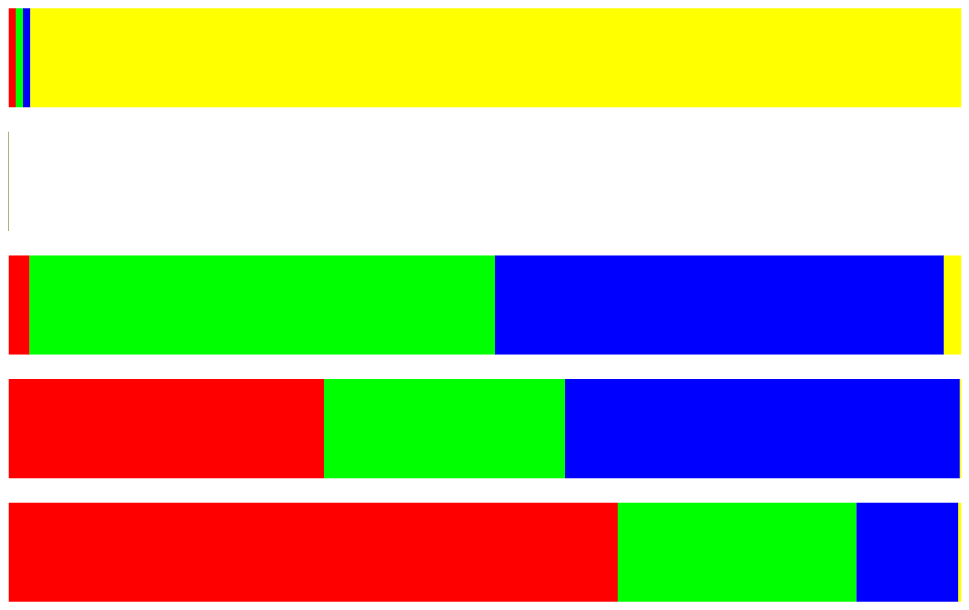,width=5.cm}}
\end{center}
\vspace*{0.5cm}
\caption{{\it The flavour content in the mass eigenstates with a
representative choice for the mixing angles:
in the 2+2 scheme with $\theta_{12} = 45^\circ, \theta_{34} = 45^\circ, 
\theta_{13} = \theta_{14} = \theta_{23} = \theta_{24} = 5^\circ$ (left);
in the 3+1 scheme with $\theta_{12} = 45^\circ, \theta_{13} = 13^\circ,
\theta_{23} = 45^\circ, \theta_{14} = \theta_{24} = \theta_{34} = 5^\circ$ (right).
The different flavours are, from lightest to darkest: $\nu_s$; $\nu_\mu$; 
$\nu_e$ and $\nu_\tau$.}}
\label{fig:smirnov}
\end{figure}

\begin{itemize}
\item {\bf The 2+2 scheme} \\
In the 2+2 scheme, that is still favoured by the data, the bound on $\nu_e$
disappearance translates into an upper limit on the combination
\be
\label{bugeychooz22}
c^2_{23} \sin^2 (2 \theta_{24}) + c^4_{24} \sin^2 (2 \theta_{23}) \leq 0.2 \ ,
\ee
whereas the bound on $\nu_e$ appearance implies
\be
\label{LSND22}
10^{-3} \leq c^2_{13} c^2_{24} \sin^2 (2 \theta_{23}) \leq 10^{-2} \ .
\ee
These bounds suggest the conservative (or even ``pessimistic'') hypothesis
adopted in \cite{Donini:1999jc}: to consider the four gap-crossing angles
$\theta_{13}, \theta_{14},\theta_{23}$ and $\theta_{24}$ to be equally small
in the mass difference region $ \Delta m^2_{LSND} \in [10^{-1}, 10^1]$ \evb .
We follow here the same hypothesis: all the gap-crossing angles are small (i.e.
less than $10^\circ$), 
with the possible exception of one angle that we leave to vary in some interval. 
The remaining angles $\theta_{12}$ and $\theta_{34}$ are directly the
solar and atmospheric mixing angles in the two-family parametrization, respectively. 
The typical flavour content of the mass eigenstates in the 2+2 scheme
is presented in Fig. \ref{fig:smirnov} (left). 
\item {\bf The 3+1 scheme} \\
This scheme is only marginally allowed (a recent study \cite{Grimus:2001mn}
shows that it is compatible with the experimental data at the 99 \% CL only). 
However, it is a natural extension of the three-family model. There are four
very small allowed region in the two-family equivalent parameter space 
\cite{Barger:2000ch}: 
\begin{enumerate}
\item 
$ \Delta m^2_{34} \simeq 0.3$ \evb \, ; \, $\sin^2 (2 \theta)_{LSND} \simeq 
2 \times 10^{-2} $ \ ; 
\item 
$ \Delta m^2_{34} \simeq 0.9$ \evb \, ; \, $\sin^2 (2 \theta)_{LSND} \simeq 
2 \times 10^{-3} $ \ ;
\item 
$ \Delta m^2_{34} \simeq 1.7$ \evb \, ; \, $\sin^2 (2 \theta)_{LSND} \simeq 
1 \times 10^{-3} $ \ ;
\item 
$ \Delta m^2_{34} \simeq 6.0$ \evb \, ; \, $\sin^2 (2 \theta)_{LSND} \simeq 
2 \times 10^{-3} $ \ .
\end{enumerate}
We restrict ourselves to case 2, for simplicity. In this case, 
we get in our parametrization for the $\nu_e$ appearance mixing parameter
\be
c^4_{34} s^2_{14} \sin^2 (2 \theta_{24}) \simeq 2 \times 10^{-3} \ . 
\ee
This bound is consistent with the conservative hypothesis of equally small
gap-crossing angles $\theta_{i4}$, that will be followed in the rest of the paper. 
In the 3+1 scheme the remaining angles, $\theta_{12}, \theta_{23}$ and $\theta_{13}$
can be obtained by the combined analysis of solar and atmospheric data
in the three-family parametrization.
The typical flavour content of the mass eigenstates in the 3+1 scheme
is presented in Fig. \ref{fig:smirnov} (right). 
\end{itemize}

\subsection{Experimental Setup: The Neutrino Factory and the Detector}
\label{setup}

In the muon rest frame, the distribution of muon antineutrinos (neutrinos)
and electron neutrinos (antineutrinos) in the decay
$\mu^\pm \raw e^\pm + \nu_e (\bar \nu_e) + \bar \nu_\mu (\nu_\mu)$ is given by:
\be
\frac{d^2 N}{dx d\Omega} = \frac{1}{4\pi} [f_0 (x) \mp {\cal P}_\mu f_1 (x) 
\cos \vartheta] \, ,
\ee
where $E_\nu$ denotes the neutrino energy, $x=2 E_{\nu}/m_{\mu}$ and 
${\cal P}_\mu$ is the average muon polarization along the beam directions. 
$\vartheta$ is the angle between the neutrino momentum vector and 
the muon spin direction and $m_\mu$ is the muon mass. The positron (electron) neutrino 
flux is identical in form to that for muon neutrinos (antineutrinos), 
when the electron mass is neglected. The functions $f_0$ and $f_1$ are given in 
Table~\ref{tab:fluxes}, \cite{gaisser}.
\begin{table}[b]
\centering
\begin{tabular}{||c|c|c||}
\hline\hline
    &  $f_0(x)$  & $f_1(x)$ \\
\hline\hline
$\nu_\mu, \rm{e}$ & $2 x^2 (3-2x)$ & $2 x^2 (1-2x)$ \\
\hline  
$\nu_e$ & $12 x^2 (1-x)$ & $12 x^2 (1-x)$ \\
\hline \hline
\end{tabular}
\caption{\it 
Flux functions.}
\label{tab:fluxes}
\end{table} 
In the laboratory frame, the neutrino fluxes, boosted along the muon momentum 
vector, are given by:
\bea
\frac{ d^2 N_{\bar \nu_\mu, \nu_\mu} }{ dy d\Omega} & = & 
   \frac{ 4 n_\mu }{ \pi L^2 m_\mu^6 } \,\  E_\mu^4 y^2 \, (1 - \beta \cos \varphi) 
   \,\, \left \{ \left [ 3 m_\mu^2 - 4 E_\mu^2 y \, (1 - \beta \cos \varphi) 
             \right ] \right . \nn \\
    & & \left. \mp \, {\cal P}_\mu 
   \left [ m_\mu^2 - 4 E_\mu^2 y \, (1 - \beta \cos \varphi)
             \right ] \right \} \, , \nn\\
\frac{ d^2 N_{\nu_e,  \bar\nu_e} }{ dy d\Omega} & = & 
   \frac{ 24 n_\mu }{ \pi L^2 m_\mu^6 } \,\, E_\mu^4 y^2 \, (1 - \beta \cos \varphi) 
   \,\, \left \{ \left [ m_\mu^2 - 2 E_\mu^2 y \, (1 - \beta \cos \varphi)
             \right ] \right . \nn \\
   & & \left. \mp \, {\cal P}_\mu 
   \left [ m_\mu^2 - 2 E_\mu^2 y \, (1 - \beta \cos \varphi)
             \right ] \right \} \, .
\label{fluxes}
\eea
Here, $\beta = \sqrt{1-m^2_\mu/E^2_\mu}$, $E_\mu$ is the parent muon energy, 
$y = E_\nu/E_\mu$, $n_\mu$ is the number of useful muons per year 
obtained from the storage ring and $L$ is the distance to the detector. 
$\varphi$ is the angle between the beam axis and the direction pointing towards 
the detector. 
We shall consider in what follows as a ``reference set-up'' a neutrino beam 
resulting from the decay of $n_\mu = 2 \times 10^{20}$ unpolarized 
positive and/or negative muons in one of the straight sections of a muon storage ring 
(i.e. we do not consider two baselines operating at the same time) per year. 
The collected muons have energy $E_\mu$ in the range $10 - 50$ GeV. 
This energy range is under discussion as a convenient goal (a definite answer 
on which is the optimal energy to run the Neutrino Factory is still missing). 
The angular divergence $\delta \varphi$ is taken to be constant, 
$\delta \varphi \sim 0.1$ mr. 

The charged current neutrino and antineutrino interaction rates can be computed
using the approximate expressions for the neutrino-nucleon cross sections
on an isoscalar target\footnote{For the $\nu_\tau$-nucleon interaction, we used 
the exact expression for the cross section taking into account the $\tau$-mass.}, 
\be
\sigma_{\nu N} \sim 0.67 \times 10^{-42} \times \frac{E_\nu}{GeV} \times m^2
\qquad
\sigma_{\bar \nu N} \sim 0.34 \times 10^{-42} \times \frac{E_\nu}{GeV} \times m^2 \ .
\ee

To explore the whole CP-conserving parameter space we need a detector
with $\tau$ tracking and ($\mu, \tau$) charge identification capability. 
As the dominant signals are expected to peak at $L/E_\nu \sim 1/\Delta m^2_{LSND}$, 
most of the parameter space can be explored in short baseline experiments (SLB), 
with $L \sim 1$ Km. At such a short distance from the source the neutrino flux
is so intense that a small detector is well suited to study CP-conserving transitions. 
In what follows we consider an hypothetical 1 ton detector, with no detailed 
calculation of background and efficiencies as a function of the neutrino energy. 
We consider a constant background $B$ at the level of $10^{-5}$ of the expected number of
charged current events, $N_{CC}$, and a constant reconstruction efficiency 
$\epsilon_\mu = 0.5$ for $\mu^\pm$ and $\epsilon_\tau = 0.35$ for $\tau^\pm$. 
The number of expected charged leptons in absence of oscillation is 
$N_{\mu^-} = 9.3 \times 10^8$ and $N_{e^+} = 4.0 \times 10^8$ for a $\mu^-$ beam
( $N_{\mu^+} = 4.7 \times 10^8$ and $N_{e^-} = 7.9 \times 10^8$ for a $\mu^+$ beam).
We also applied a conservative cut on the neutrino energy: neutrinos with
$E_\nu \leq 5$\gev have not been included in our results.

To extend our analysis to the CP-violating parameter space, a larger detector must 
be considered: we choose an hypothetical $10$ Kton detector, located a bit farther 
from the neutrino source, at $L = O(10-100)$ Km. 
Also in this case $\mu$ and $\tau$ charge identification 
capability is needed, being the $\nu_\mu \to \nu_\tau$ transitions the optimal
channel to observe CP-violation in a four-family model (as will be explained
in the following). The same background and reconstruction efficiencies
as for the CP-conserving sector are included, and again a fiducial cut 
on neutrinos with $E_\nu \leq 5$\gev is applied.

\section{Sensitivity reach of the Neutrino Factory}
\label{sensi}

We concentrate now on the sensitivity to the different gap-crossing angles
that appear in the oscillation probabilities when only the LSND mass difference
is taken into account, namely eqs. (\ref{eq:emu22}-\ref{eq:ee22}) for the 2+2 scheme 
and eqs. (\ref{eq:emu31}-\ref{eq:ee31}) for the 3+1 scheme\footnote{
Although we refer to the one- or two-mass dominance approximation (in the 
section devoted to CP-violating observables) formulae, all the numerical results have been
obtained with the exact expressions for the transition probabilities.}.

We define the sensitivity in the appearance channel as follows: the number of
total expected events for a given flavour $\nu_\alpha$ is
\be
N_{tot} = N_\alpha^B \pm \Delta N_\alpha^B + N_\beta
\ee
where
\bea
N_\alpha^B &=& N_\alpha \cdot B \, , \\
N_\beta &=& N_\alpha \, <P (\nu_\alpha \to \nu_\beta )> \, , 
\eea
with $N_\alpha$ the number of expected events in the absence of oscillation, 
$B$ the fractional background (we consider $B = 10^{-5}$) 
and $<P (\nu_\alpha \to \nu_\beta )>$ the transition probability averaged 
over the $\nu_\alpha$ flux and the CC interaction cross-section. 
Fluctuations over the background are taken to be gaussian, 
$\Delta N_\alpha^B = \sqrt{ N_\alpha^B }$. The excluded zone at 90 \% CL 
(following \cite{Feldman:1998qc}) if no event is observed is the region to the right of 
\be
N_\beta = 1.65 \; \Delta N_\alpha^B \, .
\ee
The sensitivity in the disappearance channel is defined as follows: 
the number of total expected events for a given flavour $\nu_\alpha$ is
\be
N_{tot} = N_\alpha \cdot ( 1 - B ) \pm \Delta [ N_\alpha \cdot ( 1 - B ) ] - N_\beta
\ee
where 
\be
N_\beta = \sum_{\beta \neq \alpha} N_\alpha \, < P (\nu_\alpha \to \nu_\beta ) >
\ee
summing over all flavours distinct from $\nu_\alpha$. In this case we compare
$N_\beta$ with the gaussian fluctuation over $N_\alpha \times (1 - B)$
(we notice that a background $B$ at the level of $10^{-5}$ plays a marginal role, 
with respect to the appearance case). Again, following \cite{Feldman:1998qc},
if no event is observed the region to the right of 
\be
N_\beta = 1.65 \; \Delta [ N_\alpha \cdot (1 - B) ] 
\ee
is excluded at 90 \% CL. 

\subsection{Sensitivity in the 2+2 scheme}
\label{sensi:2+2}

We recall here the results of \cite{Donini:1999jc, Donini:2000he, Donini:2000ma},
albeit rederived with slightly different input parameters.
In the one-mass dominance approximation, the CP-conserving parameter space
consists of four rotation angles ($\theta_{13}, \theta_{14}, \theta_{23}$ and 
$\theta_{24}$) and one phase, $\delta_3$. In what follows we set $\delta_3 = 0$. 
The useful channels to measure or put severe upper limits on the gap-crossing
angles at the Neutrino Factory are the following (for a $\mu^-$ decay): 
\bea
  \bar{\nu}_e & \to  \bar{\nu}_\mu & \to  \mu^+
                             \qquad (\mu^+ \, {\rm appearance}) \nn \\
  \nu_\mu & \to  \nu_\mu & \to  \mu^- 
                             \qquad (\mu^- \, {\rm disappearance}) \nn \\
  \bar{\nu}_e & \to  \bar{\nu}_\tau & \to  \tau^+ 
                             \qquad (\tau^+ \, {\rm appearance}) \nn \\
  \nu_\mu & \to  \nu_\tau & \to  \tau^- 
                             \qquad (\tau^- \, {\rm appearance}). 
\eea

In order to present the sensitivity to a specific $\sin^2 \theta$, we adopt
the following approach: we vary $\sin^2 \theta$ between $10^{-7}$ and 1; 
the remaining three angles are considered to be already known: 
two of them are fixed to a small value, $\theta_{ij} = 2^\circ$, 
and the third one is varied from $1^\circ$ to $60^\circ$. 
The remaining parameters (those measured in solar and atmospheric experiments) are
taken as follows: 
\bea
\theta_{12} &=& 45^\circ \, , \, \theta_{34} = 45^\circ \, ; \nn \\
\Delta m^2_{12} &=& 10^{-4} {\rm \ev}^2 \, , \, 
\Delta m^2_{34} = 3.5 \times 10^{-3} {\rm \ev}^2 \, . \nn
\eea 
The large mass difference $\Delta m^2_{23}$ is varied from $10^{-3}$ to $10^2$ \evb.
At $L = 1$ Km matter effects are not relevant, since such a baseline is short
enough to be completely above ground. We consider $2 \times 10^{20}$ useful
muons per year and 5 years of data taking, with $E_\mu = 20$ \gev.
For simplicity, the Neutrino Factory
is supposed to be working with negative muons only.

\begin{itemize}
\item {\bf Sensitivity to $\sin^2 \theta_{23}$: $\mu^+$ appearance}
\end{itemize}

\begin{figure}[t]
\vspace{0.1cm}
\centerline{
\epsfig{figure=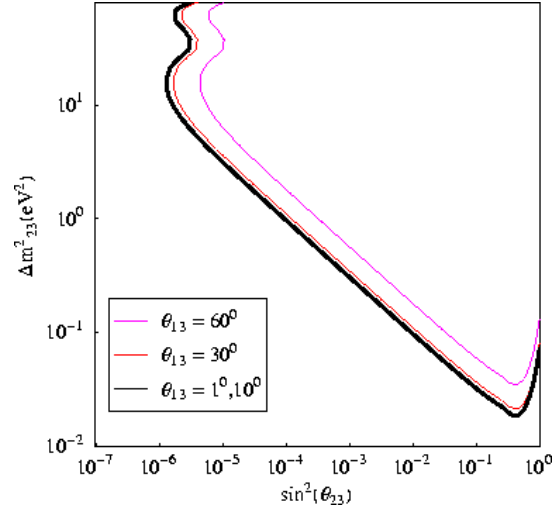,height=7.6cm,angle=0}}
\caption{\it{Sensitivity reach in the $\sin^2 \theta_{23} / \Delta m_{23}^2$ plane 
at different values of $\theta_{13}= 1^\circ, 10^\circ, 30^\circ$ and $60^\circ$
for $\mu^+$ appearance in the 2+2 scheme.}} 
\label{fig:mu23app22}
\end{figure}

The $\mu^+$ appearance channel (the so-called ``wrong-sign'' muons)
is particularly sensitive to $\theta_{23}$. 
Fig. \ref{fig:mu23app22} shows the 90 \% CL exclusion curve
in the $\sin^2 \theta_{23} / \Delta m_{23}^2$ plane for different
values of $\theta_{13}$. In the LSND-allowed region 
($\Delta m^2_{23}$ in the $10^{-1} - 10^1$ \evb range) the dependence 
on $\theta_{13}$ is mild: $\sin^2 \theta_{23}$ can reach $10^{-6}$ for $\theta_{13} 
\simeq 1^\circ$ or $6 \times 10^{-6}$ for $\theta_{13} \simeq 60^\circ$. 

\begin{itemize}
\item {\bf Sensitivity to $\sin^2 \theta_{13}$: $\mu^-$ disappearance}
\end{itemize}

\begin{figure}[t]
\vspace{0.1cm}
\centerline{
\epsfig{figure=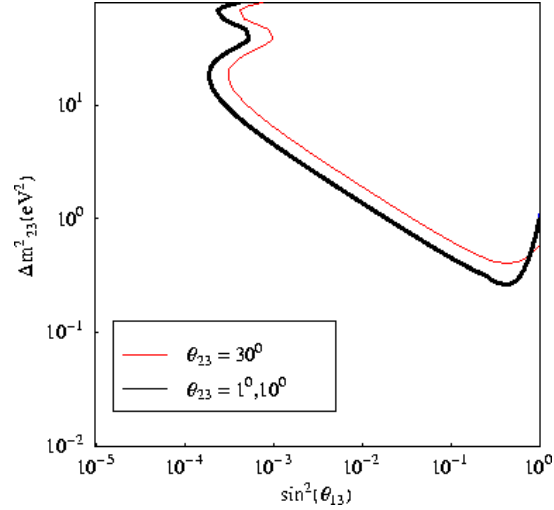,height=7.6cm,angle=0}}
\caption{\it{Sensitivity reach in the $\sin^2 \theta_{13} / \Delta m_{23}^2$ plane 
at different values of $\theta_{23}= 1^\circ, 10^\circ$ and $30^\circ$
for $\mu^-$ disappearance in the 2+2 scheme.}} 
\label{fig:mu13dis22}
\end{figure}

In Fig. \ref{fig:mu13dis22} we present the 90 \% CL exclusion curve in the 
$\sin^2 \theta_{13} / \Delta m_{23}^2$ plane, at different values of 
$\theta_{23}= 1^\circ, 10^\circ$ and $30^\circ$, for the $\mu^-$ disappearance
channel. In \cite{Donini:1999jc} it was observed that this channel proves more sensitive 
to $\sin^2 \theta_{13}$ than the $\mu^+$ appearance one for small values of 
$\theta_{23}$. 
On the contrary, the $\mu^+$ appearance channel has the larger sensitivity 
attained for large values of $\theta_{23}$, a scenario somewhat disfavoured by 
the LSND measurement. In the $\mu^-$ disappearance channel, the Neutrino Factory 
can put an upper bound to $\sin^2 \theta_{13}$ at the $10^{-4}-10^{-2}$ level for 
$\Delta m^2_{23}$ in the $10^{-1} - 10^1 $ \evb range.

\begin{itemize}
\item {\bf Sensitivity to $\sin^2 \theta_{14}$ and $\sin^2 \theta_{24}$: 
$\tau^-$ appearance}
\end{itemize}

\begin{figure}[t]
\vspace{0.1cm}
\centerline{
\epsfig{figure=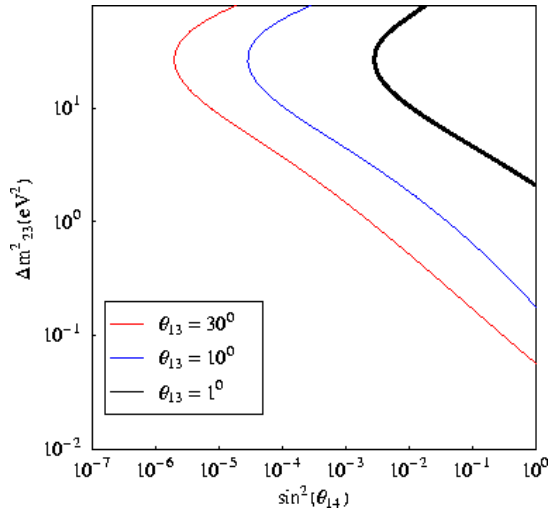,height=7.6cm,angle=0}}
\caption{\it{Sensitivity reach in the $\sin^2 \theta_{14} / \Delta m_{23}^2$ plane 
at different values of $\theta_{13}= 1^\circ, 10^\circ$ and $30^\circ$
for $\tau^-$ appearance in the 2+2 scheme.}} 
\label{fig:taum14app22}
\end{figure}

The $\tau^-$ appearance channel is quite sensitive to both $\sin^2 \theta_{14}$ and 
$\sin^2 \theta_{24}$. Fig. \ref{fig:taum14app22} illustrates the sensitivity 
to $\sin^2 \theta_{14}$ as a function of $\theta_{13}$: for about $1^\circ$, 
sensitivities of the order of $10^{-2}$ are attainable, while for $10^\circ$ 
values as small as $4 \times 10^{-5}$ can be reached. For even larger values of 
$\theta_{13}$ it goes down to $10^{-6}$ (we recall that $\theta_{13}$ 
is not severely constrained by the Bugey-CHOOZ experimental bounds, 
eq. (\ref{bugeychooz22})).

\begin{figure}[t]
\vspace{0.1cm}
\centerline{
\epsfig{figure=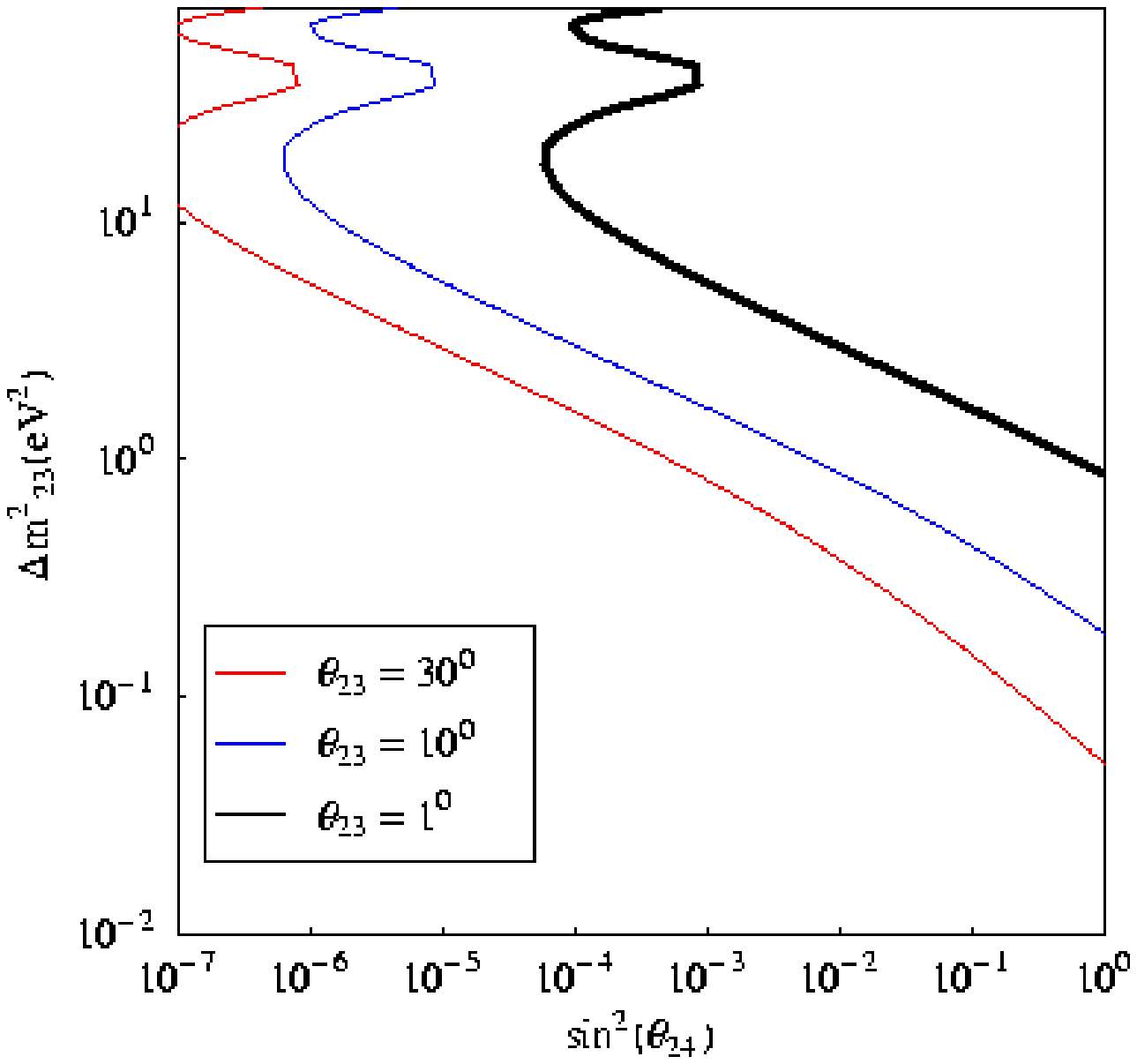,height=7.6cm,angle=0}}
\caption{\it{Sensitivity reach in the $\sin^2 \theta_{24} / \Delta m_{23}^2$ plane 
at different values of $\theta_{23}= 1^\circ, 10^\circ$ and $30^\circ$
for $\tau^-$ appearance in the 2+2 scheme.}} 
\label{fig:taum24app22}
\end{figure}

Fig. \ref{fig:taum24app22} depicts the foreseeable sensitivity reach to 
$\sin^2 \theta_{24}$ as a function of $\theta_{23}$ : for small values of 
$\theta_{23}$ the sensitivity to $\sin^2 \theta_{24}$ attains level as low
as $10^{-6}$. 

In contrast, the $\tau^+$ appearance channel looks less promising, 
for $\delta_3 = 0$. Due to the relative negative sign between the two terms in the 
analytic expression for $P(\nu_e \to \nu_\tau)$, eq.~(\ref{eq:etau22}),
cancellations for particular values of the angles occur, resulting
in a decreasing sensitivity in specific regions of the parameter space.
This sensitivity suppression is absent in the $\tau^-$ channel as the
relative sign between the two terms in $P(\nu_\mu \to \nu_\tau)$, 
eq.~(\ref{eq:mutau22}), is positive\footnote{
The same argument holds, albeit interchanging the two channels, 
for $\delta_3 = \pi$.}.

\subsection{Sensitivity in the 3+1 scheme}
\label{sensi:3+1}

In the one-mass dominance approximation, the CP-conserving parameter space
consists of three rotation angles ($\theta_{14}, \theta_{24}$ and $\theta_{34}$)
and no phases. The useful channels to measure or put severe upper limits on the 
gap-crossing angles at the Neutrino Factory are (for a $\mu^-$ decay): 
\bea
  \bar{\nu}_e & \to  \bar{\nu}_\mu & \to  \mu^+
                             \qquad (\mu^+ \, {\rm appearance}) \nn \\
  \nu_\mu & \to  \nu_\tau & \to  \tau^- 
                             \qquad (\tau^- \, {\rm appearance}). 
\eea
We will see in the following that these two channels are optimal to study
the whole CP-conserving 3+1 parameter space.

We adopt the same approach as for the 2+2 scheme: 
we vary $\sin^2 \theta$ between $10^{-7}$ and 1; the other two angles
are considered to be already known: one of them is fixed to a small value, 
$\theta_{ij} = 2^\circ$, and the second one is varied from $1^\circ$ to $60^\circ$. 
The remaining parameters (those measured in solar and atmospheric experiments) are 
taken, following \cite{Fogli:1999yz}, as: 
\bea
\theta_{12} &=& 22.5^\circ \, , \, \theta_{13} = 13^\circ \, , \, 
\theta_{23} = 45^\circ \, ; \nn \\
\Delta m^2_{12} &=& 10^{-4} {\rm \ev}^2 \, , \, 
\Delta m^2_{23} = 3.5 \times 10^{-3} {\rm \ev}^2 \, . \nn
\eea
We consider $2 \times 10^{20}$ useful
muons per year and 5 years of data taking, with $E_\mu = 20$ \gev.

\begin{itemize}
\item {\bf Sensitivity to $\sin^2 \theta_{14}$ and $\sin^2 \theta_{24}$: 
$\mu^+$ appearance}
\end{itemize}

\begin{figure}
\begin{center}
\begin{tabular}{cc}
\epsfxsize6.5cm\epsffile{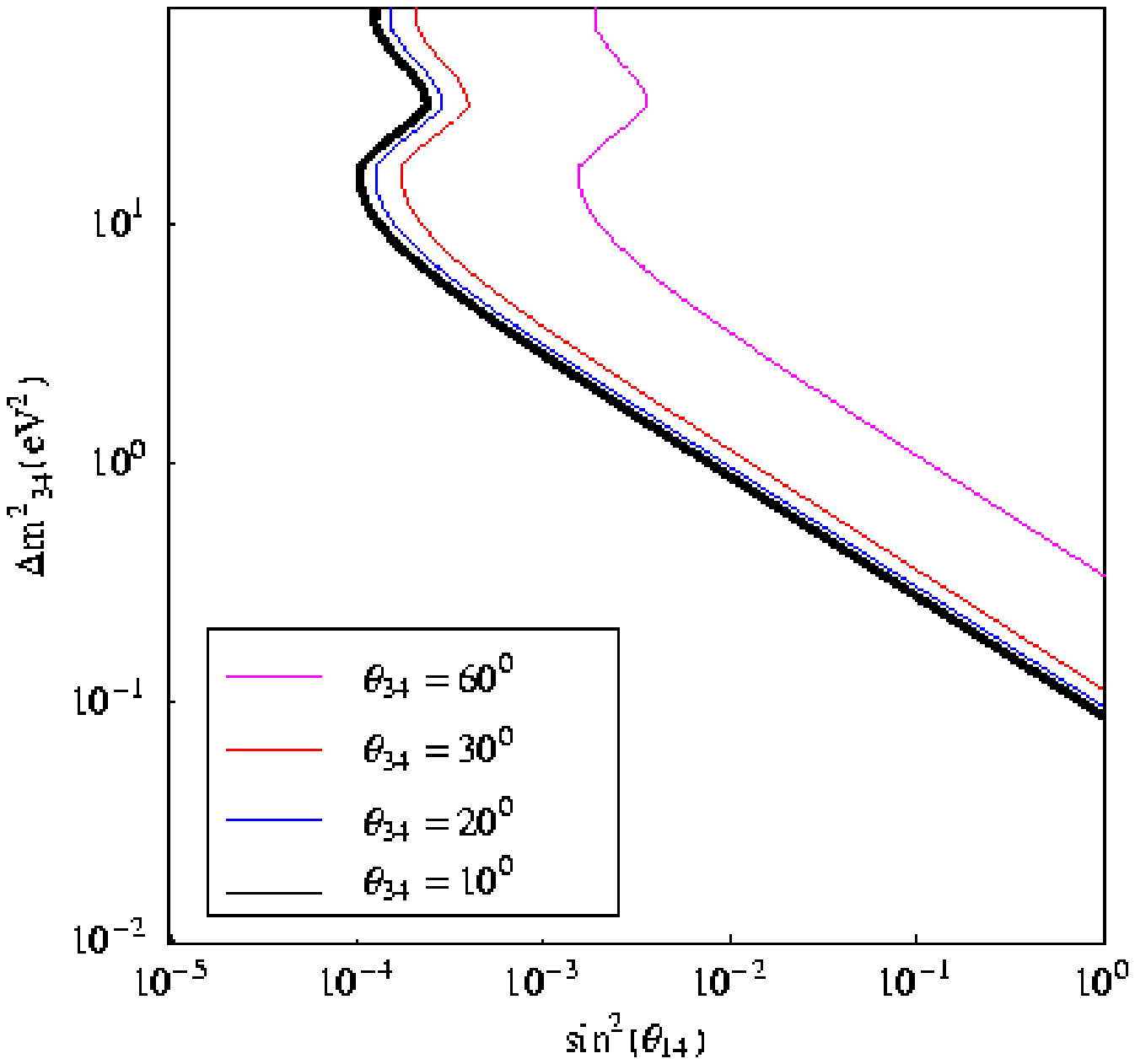} & 
\epsfxsize6.5cm\epsffile{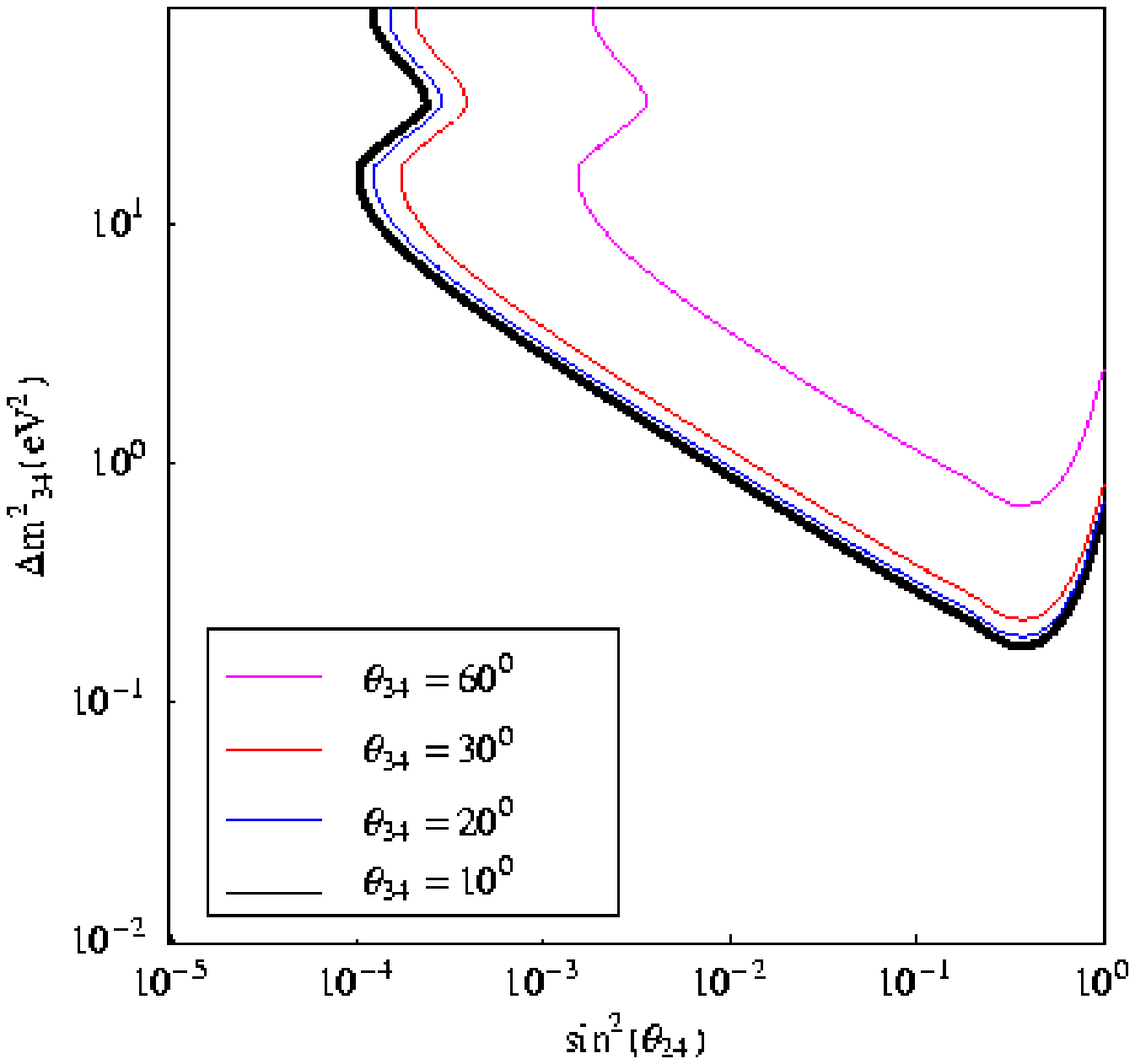}  \\
\end{tabular}
\caption{\it{Sensitivity reach 
in the $\sin^2 \theta_{14} / \Delta m_{34}^2$ plane (left) and
in the $\sin^2 \theta_{24} / \Delta m_{34}^2$ plane (right) 
at different values of $\theta_{34}= 1^\circ, 10^\circ, 30^\circ$ and $60^\circ$
for $\mu^+$ appearance in the 3+1 scheme.}} 
\label{fig:mu1424app31}
\end{center}
\end{figure}

The $\mu^+$ appearance channel is particularly sensitive to both $\sin^2 \theta_{14}$
and $\sin^2 \theta_{24}$. 
Fig. \ref{fig:mu1424app31} shows the 90 \% CL exclusion curve
in the $\sin^2 \theta_{14} / \Delta m_{34}^2$ plane (left) and
in the $\sin^2 \theta_{24} / \Delta m_{34}^2$ plane (right) for different
values of $\theta_{34}$. The dependence on $\theta_{34}$ is very mild
for small values of $\theta_{34}$.
In the LSND-allowed region, $\Delta m^2_{34} \in [10^{-1},10^1]$ \evb, 
both $\sin^2 \theta_{14}$ and $\sin^2 \theta_{24}$
can reach $10^{-4}$ for $\theta_{34} \leq 30^\circ$ or $10^{-3}$ for 
$\theta_{34} \simeq 60^\circ$. 

\begin{figure}[h!]
\begin{center}
\epsfxsize8.5cm\epsffile{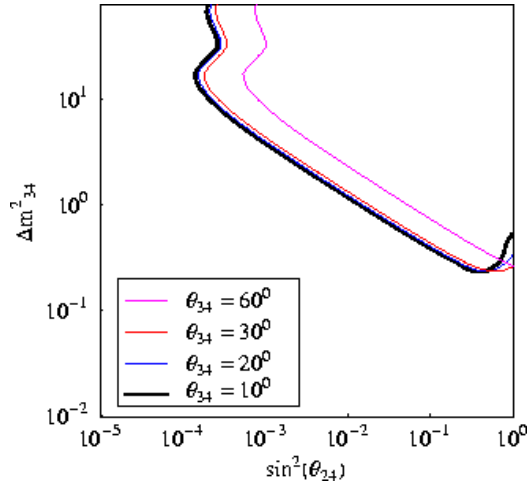}
\vspace{-1.5truecm}
\hspace{-2.0truecm}
\caption{\it{Sensitivity reach 
in the $\sin^2 \theta_{24} / \Delta m_{34}^2$ plane
at different values of $\theta_{34}= 1^\circ, 10^\circ, 30^\circ$ and $60^\circ$
for $\mu^-$ disappearance in the 3+1 scheme.}} 
\label{fig:mu24dis31}
\end{center}
\end{figure}

The $\mu^-$ disappearance channel is not sensitive to $\theta_{14}$, but
can explore approximatively the same region as the appearance channel 
in the $\sin^2 \theta_{24} / \Delta m_{34}^2$ plane, see Fig.~\ref{fig:mu24dis31}.

\begin{itemize}
\item {\bf Sensitivity to $\sin^2 \theta_{34}$: $\tau$ appearance}
\end{itemize}

\begin{figure}
\begin{center}
\begin{tabular}{cc}
\epsfxsize7cm\epsffile{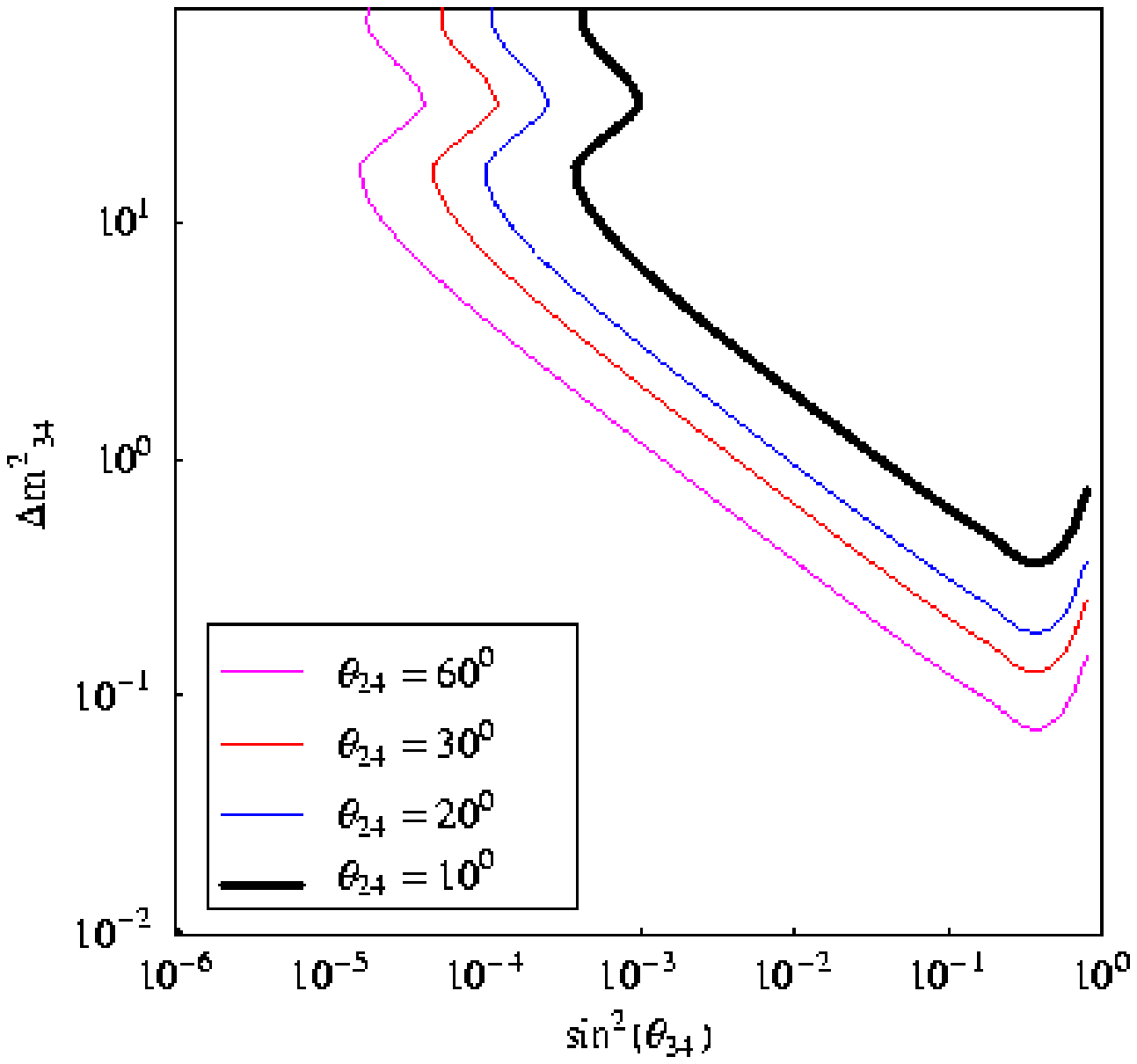} & 
\epsfxsize7cm\epsffile{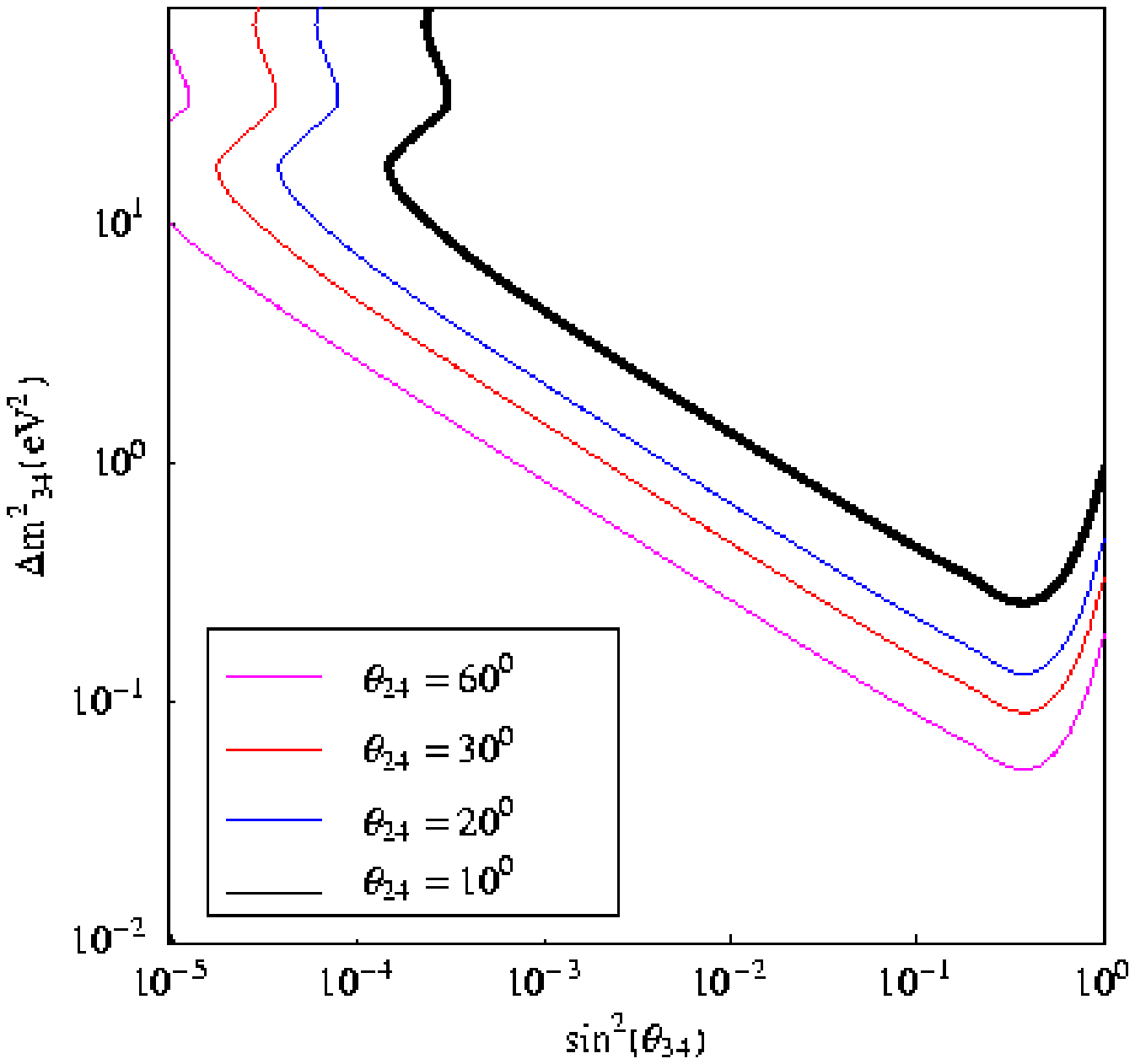}  \\
\end{tabular}
\caption{\it{Sensitivity reach 
in the $\sin^2 \theta_{34} / \Delta m_{34}^2$ plane 
at different values of $\theta_{14}= 1^\circ, 10^\circ, 30^\circ$ and $60^\circ$ 
for $\tau^+$ appearance in the 3+1 scheme (left) and
at different values of $\theta_{24}= 1^\circ, 10^\circ, 30^\circ$ and $60^\circ$ 
for $\tau^-$ appearance in the 3+1 scheme (right).}} 
\label{fig:tau34app31}
\end{center}
\end{figure}

Both $\tau$ appearance channel are equally sensitive to $\sin^2 \theta_{34}$, 
as can be seen in eqs. (\ref{eq:etau31}) and (\ref{eq:mutau31}).
Fig. \ref{fig:tau34app31} shows the 90 \% CL exclusion curve
in the $\sin^2 \theta_{34} / \Delta m_{34}^2$ plane for $\tau^+$ appearance
at different values of $\theta_{14}$ (left) and
for $\tau^-$ appearance at different values of $\theta_{24}$ (right).
In the LSND-allowed region, $\sin^2 \theta_{34}$
can reach some units in $10^{-5}$ for $\theta_{14}, \theta_{24} \leq 30^\circ$ 
or $10^{-5}$ for $\theta_{14}, \theta_{24} \simeq 60^\circ$. 

\subsection{Remarks and conclusions on the sensitivity reach} 
\label{sensi:concl}

The results of the previous subsections show that a Neutrino Factory with 
$n_\mu = 2 \times 10^{20}$ useful muons per year and a small detector 
of $O(1)$ ton size with $\tau$ tracking and $(\mu, \tau)$ charge identification
capability can severely constrain the whole four-family model 
CP-conserving parameter space, both in the 2+2 scheme and 3+1 scheme.
In the former, the sensitivity reach to all gap-crossing angles in the 
LSND-allowed region is at the level of $\sin^2 \theta \geq 10^{-6} - 10^{-4}$, 
depending on the specific angle considered. In the latter 
the sensitivity reach is at the level of $\sin^2 \theta \geq 10^{-5} - 10^{-3}$, 
depending on the specific angle considered, slightly less than in the 2+2 case.

This results can be easily understood in terms of a simple power counting argument. 
Consider the gap-crossing angles equally small, 
$\sin \theta_{ij} \simeq \epsilon$ (with $\theta_{ij}$ some gap-crossing angle 
in the 2+2 or the 3+1 scheme). In the 2+2 scheme, the 
CP-conserving transition probabilities become: 
\bea
\PCPCDUE(\nu_e \to \nu_\mu) &=& 
4 \epsilon^2 \,
\sin^2 \left ( \frac{\Delta m_{23}^2 L}{4 E} \right ) + O(\epsilon^4) \ , \nn \\
\PCPCDUE(\nu_e \to \nu_\tau) & = & 
4 \epsilon^2 \,
\sin^2 \left ( \frac{\Delta m_{23}^2 L}{4 E} \right ) + O(\epsilon^4) \ , \nn \\
\PCPCDUE(\nu_\mu \to \nu_\tau) & = & O(\epsilon^4) \ , \nn \\
\PCPCDUE(\nu_\mu \to \nu_\mu) & = & \PCPCDUE(\nu_e \to \nu_e) =
1 - 8 \epsilon^2 
\sin^2 \left ( \frac{\Delta m_{23}^2 L}{4 E} \right ) + O(\epsilon^4) \ . \nn
\eea
We notice that, with the exception of $\nu_\mu \to \nu_\tau$, the 
transition probabilities are generically of $O(\epsilon^2)$. 
In the 3+1 scheme, on the contrary, 
\bea
\PCPCTRE(\nu_e \to \nu_\mu) &=& \PCPCTRE(\nu_e \to \nu_\tau) = 
\PCPCTRE(\nu_\mu \to \nu_\tau) = O(\epsilon^4) \ , \nn \\
\PCPCTRE(\nu_\mu \to \nu_\mu) & = & \PCPCTRE(\nu_e \to \nu_e) = 
1 - 4 \epsilon^2 
\sin^2 \left ( \frac{\Delta m_{34}^2 L}{4 E} \right ) + O(\epsilon^4) \ ; \nn
\eea
all the appearance transition probabilities are generically of $O (\epsilon^4)$. 
This explains the (slight) decrease in the sensitivity in the 3+1 scheme
with respect to the 2+2 scheme. The 2+2 $\nu_\mu \to \nu_\tau$ case
is similar to the generic situation in the 3+1 scheme: Fig. \ref{fig:taum24app22}
and Fig. \ref{fig:tau34app31} (right) show the same sensitivity reach, indeed. 

Finally, we also present the MNS mixing matrix in the two schemes at
$O(\epsilon)$: 
\bea
U^{2+2}=
\left(
\begin{array}{cccc}
         1 &                          0 & \epsilon                & \epsilon \\ 
         0 &                          1 & \epsilon \, e^{i \delta_3} & \epsilon \\ 
- \epsilon & - \epsilon \, e^{-i \delta_3} &                       1 &        0 \\
- \epsilon & - \epsilon                 &                       0 &        1 
\end{array}
\right) + O ( \epsilon^2) \ , 
\label{eq:matrix22} \\
\nn \\
U^{3+1}=
\left(
\begin{array}{cccc}
         1 &          0 &          0 & \epsilon \\ 
         0 &          1 &          0 & \epsilon \\ 
         0 &          0 &          1 & \epsilon \\
- \epsilon & - \epsilon & - \epsilon &        1 
\end{array}
\right) + O (\epsilon^2) \ .
\label{eq:matrix31}
\eea
We remind that in the 2+2 scheme the sterile neutrino is in the first row, 
$\nu_\alpha = \{ \nu_s, \nu_e, \nu_\mu, \nu_\tau \}$,
whereas in the 3+1 scheme is in the last one, 
$\nu_\alpha = \{ \nu_e, \nu_\mu, \nu_\tau, \nu_s \}$. 
We can build the main contributions to the transition probabilities 
first writing $\nu_\alpha, \nu_\beta$ as linear
combinations of mass eigenstates with coefficients given in 
eqs. (\ref{eq:matrix22}) and (\ref{eq:matrix31}), and then computing 
$| \langle \nu_\alpha (t) | \nu_\beta \rangle |^2$. In this way it is simple
to derive the behaviour of all the transition probabilities
$P (\nu_\alpha \to \nu_\beta)$.

\section{CP-violating Observables}
\label{cp-viola}

Genuine CP-violating effects manifest only when at least two mass differences
are simultaneously non-vanishing. In the three-family model, the CP-violating
contribution to the oscillation probabilities can be written as \cite{Krastev:1988yu}:
\be
\PCPV = \pm 2 J \left( \sin 2 \Delta_{12} 
+ \sin 2 \Delta_{23} - \sin 2 \Delta_{13} \right) \nn
\ee
with $J = c_{12} s_{12} c^2_{13} s_{13} c_{23} s_{23} \sin \delta$ the Jarlskog
factor and $\Delta_{ij}$ as defined in eq. (\ref{def:deltaij})
(the $\pm$ sign refers to neutrinos/antineutrinos). 
If $\Delta_{12} \ll \Delta_{23}$, $\PCPV$ is negligible.
Therefore, for three-family neutrino mixing the size of the CP-violating
oscillation probability depends on the range of $\Delta m^2_{12}$, the solar mass difference.
In \cite{Cervera:2000kp,Donini:2000ky} it has been shown that a maximal
phase, $|\delta| = 90^\circ$, can be measured at 90\% CL 
if the LMA-MSW solution with $\Delta m^2_{12} \ge 2 \times 10^{-5}$ \evb
is considered. For smaller values of the solar mass difference, 
it seems impossible to measure $\delta$ with the foreseeable beams. 
However, in the four-family model the situation is totally different \cite{Donini:2001hw}: 
we can consider CP-violating observables that do not depend on $\Delta_{sol}$, 
but on $\Delta_{atm}$ and $\Delta_{LSND}$ only. Therefore, for four-family neutrino
mixing, large CP-violating effects are possible (depending on the specific value of the 
phases $\delta_i$). In the two-mass dominance approximation the parameter
space consists of five rotation angles and two phases, both for the 2+2 and 3+1 schemes.
In the 2+2 scheme, the CP-violating oscillation probabilities are given by 
eqs. (\ref{eq:emu22cp}-\ref{eq:mutau22cp}). We notice that, in these expressions, 
the size of the CP-violating probability is linearly dependent on the atmospheric
mass difference, $\Delta_{34}$, whereas the location of the maximum depends on 
the LSND mass difference, $\Delta_{23}$. Therefore, we expect a maximum in the 
CP-violating observable at $O(10)$ Km for neutrinos of $E_\nu = O(10)$ GeV. 
With such a short baseline, matter effects are completely negligible. 
In the 3+1 scheme, eqs. (\ref{eq:emu31cp}-\ref{eq:mutau31cp}),
similar results are obtained for $\Delta_{atm} = \Delta_{23}$ 
and $\Delta_{LSND} = \Delta_{34}$. This has to be compared with the three-family model, 
where the size of the CP-violating probability depends linearly on $\Delta_{sol}$
and the maximum location depends on $\Delta_{atm}$: in this case, the maximum
of the CP-observable is expected at $O (1000)$ Km, and therefore matter effects
are extremely important \cite{Freund:2000gy}. 

CP-odd effects are observable in appearance channels, while 
disappearance ones are only sensitive to the CP-even part. 
The easiest way to measure CP-violation in oscillation is to build a 
CP-asymmetry or a T-asymmetry \cite{DeRujula:1999hd}:
\bea
A_{\alpha \beta}^{CP} & \equiv &
\frac{ P(\nu_\alpha \raw \nu_\beta) - P(\bar{\nu}_\alpha \raw \bar{\nu}_\beta)}
     { P(\nu_\alpha \raw \nu_\beta) + P(\bar{\nu}_\alpha \raw \bar{\nu}_\beta)} \ ,
\label{CPodd} \\
A_{\alpha \beta}^{T} & \equiv & 
\frac{ P(\nu_\alpha \raw \nu_\beta) - P(\nu_\beta \raw \nu_\alpha)}
     { P(\nu_\alpha \raw \nu_\beta) + P(\nu_\beta \raw \nu_\alpha)} \ .
\label{Todd} 
\eea
$A_{\alpha \beta}^{CP}$ and $A_{\alpha \beta}^{T}$ are theoretically 
equivalent in vacuum due to $CPT$, and matter effects are negligible
at the short distances under consideration.
Their extraction from data at a Neutrino factory is quite different, though.
Consider, as an example, the $(\nu_e \to \nu_\mu)$ channel. 
The CP-asymmetry, $A_{e \mu}^{CP}$, would be measured by first
extracting $P(\nu_e \raw \nu_\mu)$ from the produced (wrong-sign) $\mu^-$s 
in a beam from $\mu^+$ decay and $P(\bar\nu_e \raw \bar\nu_\mu)$ from the 
charge conjugate beam and process. Notice that even if the fluxes are very 
well known, this requires a good knowledge of the cross section ratio 
$\sigma(\bar \nu_\mu \to \mu^+)/\sigma(\nu_\mu \to \mu^-)$. Conversely, the 
measurement of the T-asymmetry, $A_{e \mu}^{T}$, requires  
to consider $P(\nu_\mu \raw \nu_e)$ and thus a good $e$ charge identification, 
that seems harder to achieve from the experimental point of view. 
In the following we will deal only with CP-asymmetries.

A central question on the observability of CP-violation is that of statistics.
We do not exploit here the explicit $E_\nu$ dependence of the CP-odd effect, 
and we consider the neutrino-energy integrated quantity: 
\begin{equation}
{\bar A}^{CP}_{ \alpha \beta} (\delta) = 
\frac{ \{ {N[l_\beta^-]}/{N_o[l_\alpha^-]} \} 
       - \{N[l_\beta^+]/N_o[l_\alpha^+]    \} }{ 
       \{  N[l_\beta^-]/N_o[l_\alpha^-]    \} 
       + \{N[l_\beta^+]/N_o[l_\alpha^+]    \} } \; ,
\label{intasy}
\end{equation}
where $l_\alpha, l_\beta$ are the charged leptons produced via CC interactions
by $\nu_\alpha, \nu_\beta$, respectively (the sign of the decaying muons is indicated 
by an upper index). $N[l_\beta^\pm]$ is the number of CC interactions 
due to oscillated neutrinos, whereas $N_o[l_\alpha^\pm]$ 
is the expected number of CC interactions in the absence of oscillations.
In order to quantify the significance of the signal, we compare the 
value of the integrated asymmetry with its error, $\Delta {\bar A}^{CP}_{ \alpha \beta}$, 
in which we include the statistical error and a conservative background estimate at
the level of $10^{-5}$.

\begin{figure}[h!]
\begin{center}
\epsfxsize9cm\epsffile{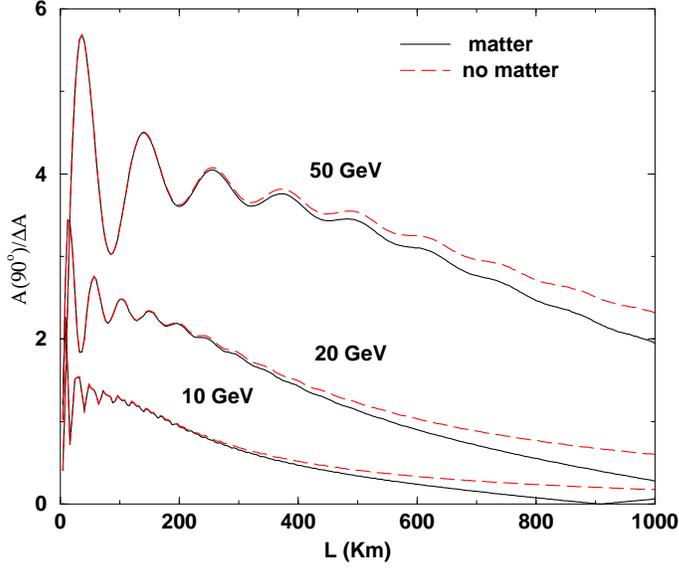} 
\caption{{\it 
Signal-to-noise ratio of the CP-violating asymmetry in the $\nu_e \to \nu_\mu$
channel with and without matter effects, for $E_\mu = 10, 20$ and $50$ GeV, 
as a function of the baseline $L$.
The parameters are: $\Delta m^2_{sol} = 10^{-4}$ \evb; 
$\Delta m^2_{atm} = 3.5 \times 10^{-3}$ \evb; $\Delta m^2_{LSND} = 1 $ \evb; 
$\theta_{12} = 22.5^\circ, \theta_{13} = 13^\circ, \theta_{23} = 45^\circ$;
$\theta_{14} = \theta_{24} = \theta_{34} = 5^\circ$; 
$\delta_2 = 0^\circ, \delta_3 = 90^\circ$. The matter parameter, 
$A = 2 E_\nu \, \sqrt{2} G_F n_e$ (with $n_e$ the electron density in the Earth)
is taken to be constant, $A = 1.1 \times 10^{-4} $ \evb / GeV. This value is 
consistent with a baseline completely contained in the Earth crust \cite{earthmodel},
true for $L \leq 4000 $ Km.}}
\label{fig:matter}
\end{center}
\end{figure}

In what follows we used the exact expression for the oscillation probabilities, 
thus including the small $\Delta_{sol}$ mass difference and the matter effects. 
The irrelevance of the latter at the considered baseline can be seen in 
Fig. \ref{fig:matter}, where 
${\bar A}^{CP}_{ e \mu} (\delta = 90^\circ) / \Delta {\bar A}^{CP}_{ e \mu}$ 
in the 3+1 scheme is presented: we see that matter effects start to be relevant 
at $O(1000)$ Km. 

Since the matter effects are negligible, the scaling laws with the muon energy $E_\mu$
and the baseline $L$ of the signal-to-noise ratio of the CP-violating asymmetry 
in eq.~(\ref{intasy}) are equivalent to those obtained in vacuum, 
\be
\frac{{\bar A}_{\alpha \beta}^{CP}}{\Delta {\bar A}_{\alpha \beta}^{CP}} 
\propto \sqrt{E_\nu}  \left| \sin \left ( \frac{\Delta m^2_{LSND} \, L}{4 E_\nu} \right ) 
\right| .
\label{eq:scaling}
\ee

\subsection{CP-violation in the 2+2 scheme}
\label{cp-viola:2+2}

We recall here the results of \cite{Donini:1999jc, Donini:2000he}, 
albeit rederived with slightly different input parameters.
In the conservative assumption of small gap-crossing angles, $\theta_{13}, \theta_{14}, 
\theta_{23}$ and $\theta_{24}$, we consider the following values for the parameters
of the MNS mixing matrix in the two-mass dominance approximation: 
\bea
\theta_{13} &=& \theta_{14} = \theta_{23} = \theta_{24} = 2^\circ \, ; \,
\theta_{12} = 45^\circ, \theta_{34} = 45^\circ \, ; \nn \\
\Delta m^2_{12} &=& 10^{-4} \ {\rm eV}^2 \, , \, 
\Delta m^2_{34} = 3.5 \times 10^{-3} \ {\rm eV}^2 \, , \, 
\Delta m^2_{23} = 1 \ {\rm eV}^2; 
\label{eq:param22cp} \\
A &= & 1.1 \times 10^{-4} \ \frac{{\rm eV}^2}{\gev} \ . \nn
\eea
The detector characteristics have been given in Sect. \ref{setup}. 

\begin{itemize}
\item {\bf Integrated asymmetry in the $\nu_e \to \nu_\mu$ channel}
\end{itemize}

\begin{figure}[h!]
\begin{center}
\epsfxsize7cm\epsffile{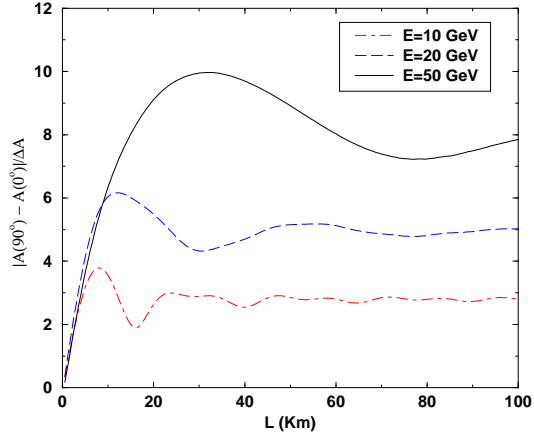} 
\caption{{\it Signal over statistical uncertainty for CP violation in the 
$\nu_e \to \nu_\mu$ channel in the 2+2 scheme, 
as a function of the baseline $L$, for three values of the
parent muon energy, $E_\mu = 10, 20$ and $50$ GeV. The parameters have been 
choosen as in eq. (\ref{eq:param22cp}), with $\delta_2 = 0^\circ, \delta_3 = 90^\circ$.}}
\label{fig:emu22cp}
\end{center}
\end{figure}

In Fig. \ref{fig:emu22cp} we show the signal-to-noise ratio of the integrated CP
asymmetry, eq. (\ref{intasy}), in the $\nu_e \to \nu_\mu$ channel, 
as a function of the distance $L$ for three values of the parent muon energy, 
$E_\mu = 10, 20$ and $50$ GeV. Matter effects, although negligible, are included. 
For this reason, we subtract to the total asymmetry ${\bar A}_{e \mu}^{CP} (90^\circ)$ 
the matter-induced asymmetry, ${\bar A}_{e \mu}^{CP} (0^\circ)$. 
We notice that a sizeable signal can be reached: for $E_\mu = 50$ GeV, approximately 10
standard deviations (sd) at $L \simeq 30$ Km can be attained. The scaling of the maximum
height with the parent muon energy follows eq. (\ref{eq:scaling}) as expected, 
increasing with $\sqrt{E_\mu}$. 

\begin{itemize}
\item {\bf Integrated asymmetry in the $\nu_e \to \nu_\tau$ channel}
\end{itemize}

\begin{figure}[h!]
\begin{center}
\epsfxsize7cm\epsffile{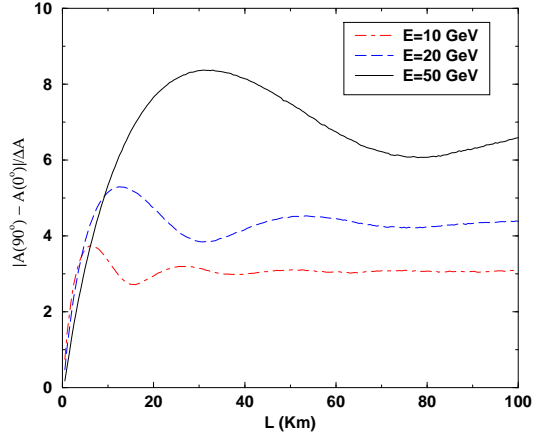} 
\caption{{\it Signal over statistical uncertainty for CP violation in the 
$\nu_e \to \nu_\tau$ channel in the 2+2 scheme, 
as a function of the baseline $L$, for three values of the
parent muon energy, $E_\mu = 10, 20$ and $50$ GeV. The parameters have been 
choosen as in eq. (\ref{eq:param22cp}), with $\delta_2 = 0^\circ, \delta_3 = 90^\circ$.}}
\label{fig:etau22cp}
\end{center}
\end{figure}

In Fig. \ref{fig:etau22cp} we show the signal-to-noise ratio of the subtracted integrated CP
asymmetry in the $\nu_e \to \nu_\tau$ channel. The results are pretty similar 
to those for the $\nu_e \to \nu_\mu$ channel, with a slightly smaller significance
at the maximum: for $E_\mu = 50$ \gev, $\sim 8$ sd can be attained at $L \simeq 30$ Km.

\begin{itemize}
\item {\bf Integrated asymmetry in the $\nu_\mu \to \nu_\tau$ channel}
\end{itemize}

\begin{figure}[h!]
\begin{center}
\epsfxsize7cm\epsffile{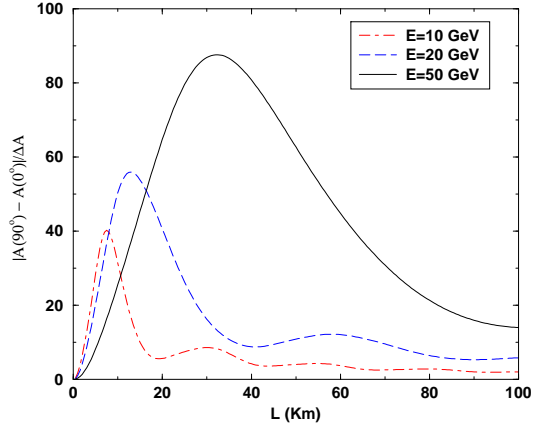} 
\caption{{\it Signal over statistical uncertainty for CP violation in the 
$\nu_\mu \to \nu_\tau$ channel in the 2+2 scheme, 
as a function of the baseline $L$, for three values of the
parent muon energy, $E_\mu = 10, 20$ and $50$ GeV. The parameters have been 
choosen as in eq. (\ref{eq:param22cp}), with $\delta_2 = 0^\circ, \delta_3 = 90^\circ$.}}
\label{fig:mutau22cp}
\end{center}
\end{figure}

In Fig. \ref{fig:mutau22cp} we show the signal-to-noise ratio of the subtracted 
integrated CP
asymmetry in the $\nu_\mu \to \nu_\tau$ channel. The results are totally
different from those relative to $\nu_e \to \nu_\mu, \nu_\tau$: 
for $E_\mu = 50$ GeV, $\sim 90$ sd can be attained at $L \simeq 30$ Km. 

\subsection{CP-violation in the 3+1 scheme}
\label{cp-viola:3+1}

In the conservative assumption of small gap-crossing angles, $\theta_{14}, 
\theta_{24}$ and $\theta_{34}$, we consider the following values for the parameters
of the MNS mixing matrix in the two-mass dominance approximation: 
\bea
\theta_{14} &=& \theta_{24} = \theta_{34} = 2^\circ \, ; \,
\theta_{12} = 22.5^\circ, \theta_{13} = 13^\circ, \theta_{23} = 45^\circ \, ; \nn \\
\Delta m^2_{12} &=& 10^{-4} \ {\rm eV}^2 \, , \, 
\Delta m^2_{23} = 3.5 \times 10^{-3} \ {\rm eV}^2 \, , \, 
\Delta m^2_{34} = 1 \ {\rm eV}^2; 
\label{eq:param31cp} \\
A &=& 1.1 \times 10^{-4} \ \frac{{\rm eV}^2}{\gev} \ . \nn
\eea

\begin{itemize}
\item {\bf Integrated asymmetry in the $\nu_e \to \nu_\mu$ channel}
\end{itemize}

\begin{figure}[h!]
\begin{center}
\epsfxsize7cm\epsffile{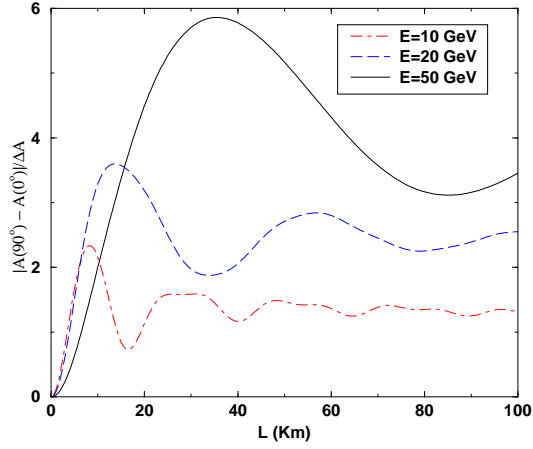} 
\caption{{\it Signal over statistical uncertainty for CP violation in the 
$\nu_e \to \nu_\mu$ channel in the 3+1 scheme, 
as a function of the baseline $L$, for three values of the
parent muon energy, $E_\mu = 10, 20$ and $50$ GeV. The parameters have been 
choosen as in eq. (\ref{eq:param31cp}), with $\delta_2 = 0^\circ, \delta_3 = 90^\circ$.}}
\label{fig:emu31cp}
\end{center}
\end{figure}

In Fig. \ref{fig:emu31cp} we show the signal-to-noise ratio of the subtracted 
integrated CP asymmetry in the $\nu_e \to \nu_\mu$ channel, 
as a function of the distance $L$ for three values of the parent muon energy, 
$E_\mu = 10, 20$ and $50$ GeV. The results are quite similar to those
obtained in the 2+2 scheme, although the significance is slightly less: 
for $E_\mu = 50$ GeV, $\sim 6$ sd at $L \simeq 40$ Km can be attained. 

\begin{figure}[h!]
\begin{center}
\epsfxsize7cm\epsffile{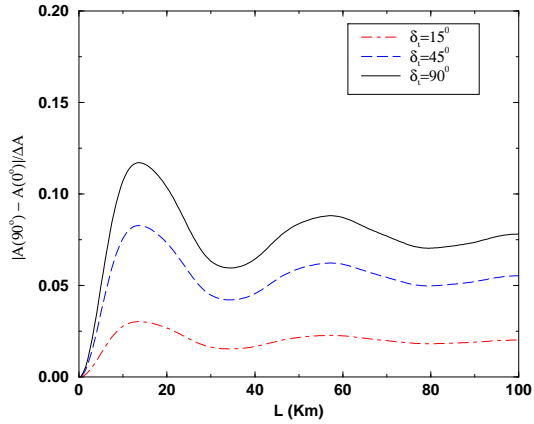} 
\caption{{\it Signal over statistical uncertainty for CP violation in the 
$\nu_e \to \nu_\mu$ channel in the 3+1 scheme, 
as a function of the baseline $L$, for three values of the
phases, $\delta_2 = \delta_3 = 15^\circ, 45^\circ$ and $90^\circ$. 
The parameters have been choosen as in eq. (\ref{eq:param31cp}), with $E_\mu = 20$ GeV.}}
\label{fig:emu31cp_delta}
\end{center}
\end{figure}

The most unfortunate case is presented in Fig. \ref{fig:emu31cp_delta}.
We show the signal-to-noise ratio of the subtracted integrated CP asymmetry
as a function of the distance $L$ for three values 
of the CP-violating phases, $\delta_2 = \delta_3 = 15^\circ, 45^\circ$ and $90^\circ$. 
In the 3+1 scheme the corresponding oscillation probability, eq.~(\ref{eq:emu31cp}), 
contains three different terms in $\sin \delta_2$, $\sin \delta_3$ (with opposite signs) 
and $\sin (\delta_2 - \delta_3)$. 
Therefore, when $\delta_2 = \delta_3$ a cancellation occurs. This is the
reason of the decrease in the significance for $\delta_2 = \delta_3 = 90^\circ$
with respect to the corresponding curve in Fig. \ref{fig:emu31cp}.
For the lowest value of the phase only $\sim 0.02$ sd can be attained. 

\newpage

\begin{itemize}
\item {\bf Integrated asymmetry in the $\nu_e \to \nu_\tau$ channel}
\end{itemize}

\begin{figure}[h!]
\begin{center}
\epsfxsize7cm\epsffile{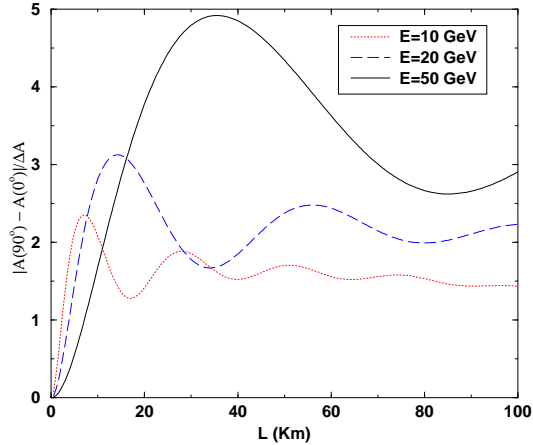} 
\caption{{\it Signal over statistical uncertainty for CP violation in the 
$\nu_e \to \nu_\tau$ channel in the 3+1 scheme, 
as a function of the baseline $L$, for three values of the
parent muon energy, $E_\mu = 10, 20$ and $50$ GeV. The parameters have been 
choosen as in eq. (\ref{eq:param31cp}), with $\delta_2 = 0^\circ, \delta_3 = 90^\circ$.}}
\label{fig:etau31cp}
\end{center}
\end{figure}

In Fig. \ref{fig:etau31cp} we show the signal-to-noise ratio of the subtracted 
integrated CP asymmetry in the $\nu_e \to \nu_\tau$ channel.
The results are pretty similar to those for the $\nu_e \to \nu_\mu$ channel, 
with a slightly smaller significance at the maximum: for $E_\mu = 50$ GeV, 
$\sim 5$ sd can be attained at $L \simeq 40$ Km.

\begin{itemize}
\item {\bf Integrated asymmetry in the $\nu_\mu \to \nu_\tau$ channel}
\end{itemize}

\begin{figure}[h!]
\begin{center}
\epsfxsize7cm\epsffile{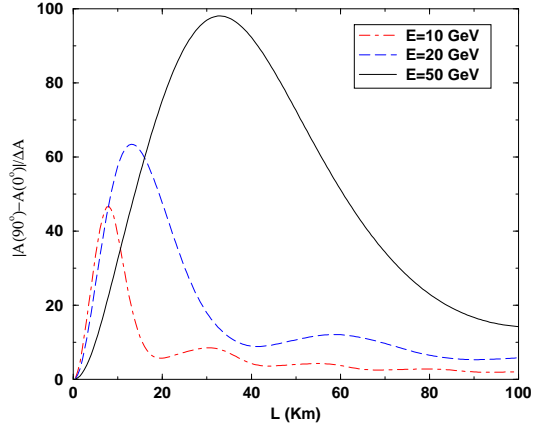} 
\caption{{\it Signal over statistical uncertainty for CP violation in the 
$\nu_\mu \to \nu_\tau$ channel in the 3+1 scheme, 
as a function of the baseline $L$, for three values of the
parent muon energy, $E_\mu = 10, 20$ and $50$ GeV. The parameters have been 
choosen as in eq. (\ref{eq:param31cp}), with $\delta_2 = 90^\circ, \delta_3 = 0^\circ$.}}
\label{fig:mutau31cp}
\end{center}
\end{figure}

In Fig. \ref{fig:mutau31cp} we show the signal-to-noise ratio of the subtracted 
integrated CP asymmetry in the $\nu_\mu \to \nu_\tau$ channel. Again, as in the 2+2
scheme, the results are totally different from those relative to 
$\nu_e \to \nu_\mu, \nu_\tau$. For $E_\mu = 50$ GeV, $\sim 100$ sd can be attained 
at $L \simeq 40$ Km. 

\begin{figure}[h!]
\begin{center}
\epsfxsize7cm\epsffile{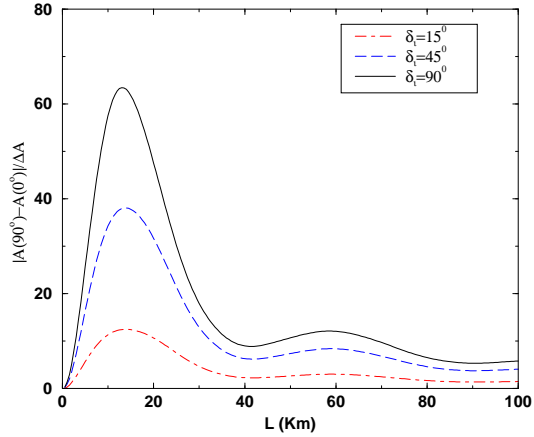} 
\caption{{\it Signal over statistical uncertainty for CP violation in the 
$\nu_\mu \to \nu_\tau$ channel in the 3+1 scheme, 
as a function of the baseline $L$, for three values of the
phases, $\delta_2 = \delta_3 = 15^\circ, 45^\circ$ and $90^\circ$. 
The parameters have been choosen as in eq. (\ref{eq:param31cp}), with $E_\mu = 20$ GeV.}}
\label{fig:mutau31cp_delta}
\end{center}
\end{figure}

In Fig. \ref{fig:mutau31cp_delta} we show the signal-to-noise ratio of the subtracted 
integrated CP asymmetry
as a function of the distance $L$ for three values of the CP-violating phases, 
$\delta_2 = \delta_3 = 15^\circ, 45^\circ$ and $90^\circ$. The oscillation 
probability at the leading order only depends on $\sin \delta_2$,
see eq. (\ref{eq:mutau31cp}). 
In this case, for the lowest value of the phase $\sim 10$ sd can still be reached. 

\subsection{Remarks and conclusions on the CP-violation observables} 
\label{cp-viola:concl}

The previous results clearly show how a maximal CP-violation is easily measurable
with a not-so-large detector of $10$ Kton size, at a baseline $L = O(10)$ Km.  
The optimal channel to observe CP-violation is the $\nu_\mu \to \nu_\tau$ channel, 
where $O (100)$ sd can be attained. Even at lower energies, $\sim 40-50$ sd
can be easily found. The number of expected $N_{\tau^\pm}$ in the detector is of $O(10^5)$. 
Notice that in the $\nu_\mu \to \nu_\tau$ channel a non-maximal 
CP-violating phase gives a significance at the level of $\sim 10$ sd even for
the lowest energy, $E_\mu = 10$ GeV. 
The other two channels, $\nu_e \to \nu_\mu, \nu_\tau$ are quite similar and give
much smaller significance. For a smaller detector mass $M$, a reduction factor
$\propto 1/ \sqrt{M}$ should be applied to all the results of this section.
Therefore, for an OPERA-like 1 Kton detector we still expect large CP-violating effects
in the $\nu_\mu \to \nu_\tau$ channel.

All of these results are obtained in both the 2+2 and the 3+1 scheme, with 
slight differences between the two. The real gain with respect to the three-family
model is that the small solar mass difference, that modules the overall size
of the CP-violating asymmetry, is traded with the much larger atmospheric mass
difference. 

In the three-family model, the largest CP asymmetry is expected in the 
$\nu_e \to \nu_\mu$ transition. We have shown that in the four-family model 
(both in the 2+2 and 3+1 scheme) this is not the case: it is 
$\nu_\mu \to \nu_\tau$ the optimal channel to study CP violation. 
The reason can be easily understood applying the power counting argument
introduced in Sect. \ref{sensi:concl}. Neglecting the background, 
the signal-to-noise ratio of the CP asymmetry 
$A_{\alpha \beta}^{CP}/ \Delta A_{\alpha \beta}^{CP}$ is: 
\be
\frac{A_{\alpha \beta}^{CP}}{\Delta A_{\alpha \beta}^{CP}} \propto 
\frac{\PCPV (\nu_\alpha \to \nu_\beta) }{\sqrt {\PCPC (\nu_\alpha \to \nu_\beta)} } 
\label{eq:powerratio}
\ee

The CP-violating transition probabilities in the approximation 
of equally small gap-crossing angles, $\sin \theta_{ij} \simeq \epsilon$ 
(with $\theta_{ij}$ a generic gap-crossing angle in the 2+2 or the 3+1 scheme), 
in the 2+2 scheme become: 
\bea
\PCPVDUE(\nu_e \to \nu_\mu) & = & 
    - 4 \epsilon^2 \sin (2 \theta_{34}) \sin (\delta_2 + \delta_3) \,
     \left( {{ \Delta m^2_{34} L }\over{4 E_\nu} } \right) \,
     \sin^2 \left (\frac{\Delta m_{23}^2 L}{4 E_\nu} \right) + O(\epsilon^4) \ , \nn \\
\label{eq:emu22cpapprox} \\
\PCPVDUE(\nu_e \to \nu_\tau) & = & 
    + 4 \epsilon^2 \sin (2 \theta_{34}) \sin (\delta_2 + \delta_3) \,
     \left( {{ \Delta m^2_{34} L }\over{4 E_\nu} } \right) \,
     \sin^2 \left (\frac{\Delta m_{23}^2 L}{4 E_\nu} \right) + O(\epsilon^4) \ , \nn \\
\label{eq:etau22cpapprox} \\
\PCPVDUE(\nu_\mu \to \nu_\tau) & = & 
    - 4 \epsilon^2 \sin (2 \theta_{34}) [ \sin \delta_2 + \sin (\delta_2 + \delta_3) ] \,
     \left( {{ \Delta m^2_{34} L }\over{4 E_\nu} } \right) \,
     \sin^2 \left (\frac{\Delta m_{23}^2 L}{4 E_\nu} \right) + O(\epsilon^4) \ ; \nn \\
\label{eq:mutau22cpapprox} 
\eea
in the 3+1 scheme become:
\bea
\PCPVTRE(\nu_e \to \nu_\mu) & = & 
     4 \epsilon^3 \sin (2 \theta_{23}) \sin (\delta_2 - \delta_3) \,
     \left( {{ \Delta m^2_{23} L }\over{4 E_\nu} } \right) \,
     \sin^2 \left (\frac{\Delta m_{34}^2 L}{4 E_\nu} \right) + O(\epsilon^4) \ , \nn \\
\label{eq:emu31cpapprox} \\
\PCPVTRE(\nu_e \to \nu_\tau) & = & 
    - 8 \epsilon^3 c_{23}^2 \sin \delta_3 \,
     \left( {{ \Delta m^2_{23} L }\over{4 E_\nu} } \right) \,
     \sin^2 \left (\frac{\Delta m_{34}^2 L}{4 E_\nu} \right) + O(\epsilon^4) \ , \nn \\
\label{eq:etau31cpapprox} \\
\PCPVTRE(\nu_\mu \to \nu_\tau) & = & 
    - 4 \epsilon^2 \sin (2 \theta_{23}) \sin \delta_2 \,
     \left( {{ \Delta m^2_{23} L }\over{4 E_\nu} } \right) \,
     \sin^2 \left (\frac{\Delta m_{34}^2 L}{4 E_\nu} \right) + O(\epsilon^4) \nn \\
\label{eq:mutau31cpapprox}
\eea
(in this case we have considered $s_{13} \sim \epsilon$ also, since 
the present experimental results in the three-family model show that 
$\theta_{13} \leq 13^\circ$ \cite{Fogli:1999yz}).

In Tab. \ref{tab:power} we report the leading order in $\epsilon$ for the 
different $\PCPC$ and $\PCPV$ in the three-family model and in both schemes
of the four-family model. In three-family, the small parameter is $s_{13} \sim \epsilon$. 
In four-family we consider equally small LSND gap-crossing angles: 
$ s_{13} = s_{14} = s_{23} = s_{24} \sim \epsilon $ for the 2+2 scheme;
$ s_{14} = s_{24} = s_{34} = \epsilon $ for the 3+1 scheme. In this 
last case we take $s_{13} \sim \epsilon$, also. We also report the leading order 
in $\epsilon$ of the signal-to-noise ratio of the various CP asymmetries, 
using eq. (\ref{eq:powerratio}). 

\begin{table}
\begin{center}
\begin{tabular}{||c||c|c|c|c||}
\hline
\hline
Scheme        & Transition & $\PCPC $ & $ \PCPV $ & $ A / \Delta A $ \\
\hline
\hline
              & $\nu_e \raw \nu_\mu $ & $\epsilon^2$ & $\epsilon$ & $ O(1) $  \\ 
Three-family  & $\nu_e \raw \nu_\tau$ & $\epsilon^2$ & $\epsilon$ & $ O(1) $  \\
            & $\nu_\mu \raw \nu_\tau$ &            1 & $\epsilon$ & $ O(\epsilon) $  \\
\hline
\hline
              & $\nu_e \raw \nu_\mu $ & $\epsilon^2$ & $\epsilon^2$ & $ O(\epsilon) $  \\ 
2+2         & $\nu_e \raw \nu_\tau$ & $\epsilon^2$ & $\epsilon^2$ & $ O(\epsilon) $  \\
            & $\nu_\mu \raw \nu_\tau$ & $\epsilon^4$ & $\epsilon^2$ & $ O(1) $  \\
\hline
\hline
              & $\nu_e \raw \nu_\mu $ & $\epsilon^4$ & $\epsilon^3$ & $ O(\epsilon) $  \\ 
3+1         & $\nu_e \raw \nu_\tau$ & $\epsilon^4$ & $\epsilon^3$ & $ O(\epsilon) $  \\
            & $\nu_\mu \raw \nu_\tau$ & $\epsilon^4$ & $\epsilon^2$ & $ O(1) $  \\
\hline
\hline
\end{tabular}
\caption{{ \it Small angles suppression in the CP-conserving 
and CP-violating oscillation probabilities, and in the signal-to-noise 
ratio of the CP asymmetries, in the three-family model and in both 
four-family model mass schemes.}}
\label{tab:power}
\end{center}
\end{table}

The last column can be easily read: in the three-family model, the $\nu_e \to \nu_\mu$
and $\nu_e \to \nu_\tau$ have a signal-to-noise ratio of the corresponding CP asymmetry
of $O(1)$ in the small angles. On the contrary, in both the 2+2 and 3+1 four-family
model, it is the $\nu_\mu \to \nu_\tau$ channel to be of $O(1)$ in the small angles, 
thus justifying {\em a posteriori} our results.
As a final remark, notice that the last column for the 2+2 and the 3+1
scheme is identical, although the corresponding CP-conserving and CP-violating
probabilities are of different order in the small angles suppression. 

\section{A magnetized iron detector with no $\tau$-tracking}
\label{fit4in4}

In this section we explore the possibility of reconstructing two angles or one
angle and a phase at a time in a short baseline experiment, $L= 40$ Km 
(the distance where the CP-violating observable of the previous section
is maximized). We focus on the $\nu_e \to \nu_\mu$ channel and consider
a realistic 10 Kton magnetized iron detector with $\mu$ charge identification 
of the type discussed in \cite{Cervera:2000vy}.
A detailed analysis of the backgrounds and detection efficiencies of this 
apparatus can be found in the literature. 
The Neutrino Factory is run with $2 \times 10^{20}$ useful muons per year for 5 
operational years for both muon polarities at $E_\mu = 50$ \gev, 
with a detector energy resolution of $\Delta E_\nu = 10$ \gev.
Our motivation is the following:
is it possible to avoid the complications connected with the $\tau$ detection
taking full advantage of the energy dependence in the $\nu_e \to \nu_\mu$ channel,
instead? 

We follow the procedure described in \cite{Cervera:2000kp}: 
let $N_{i,p}$ be the total number of wrong-sign muons detected 
when the Neutrino Factory is run in polarity  $p=\mu^+,\mu^-$, 
grouped in 5 energy bins specified by $i = $ 1 to 5.
In order to simulate a typical experimental situation we generate 
a set of ``data'' 
$n_{i,p}$ as follows: for a given value of the oscillation parameters, 
the expected number of events, $N_{i,p}$, is computed; taking into account backgrounds and 
detection efficiencies per bin, $b_{i,p}$ and $\epsilon_{i,p}$ 
(as quoted in \cite{Cervera:2000kp}), we perform a Gaussian 
(or Poisson, depending on the number of events) 
smearing to mimic the statistical uncertainty: 
\begin{eqnarray} 
n_{i,p} = \frac{ {\rm Smear} (N_{i,p} \epsilon_{i,p} + b_{i,p}) - 
b_{i,p}}{\epsilon_{i,p}} \,. 
\end{eqnarray}
The "data" are then fitted to the theoretical expectation as a function of
the mixing matrix parameters under study, using a $\chi^2$ minimization:
\be
\chi^2 = \sum_p \sum_i 
\left(\frac{ n_{i,p} \, - \, 
N_{i,p}}{\delta n_{i,p}}\right)^2 \, ,
\label{chi2}
\ee
where $\delta n_{i,p}$ is the error on $n_{i,p}$ 
(we include no error in the efficiencies).

First consider the simultaneous measurement of $\theta_{14}$ and $\theta_{24}$.
In the one-mass dominance approximation the $\nu_e \to \nu_\mu$
transition probability, eq. (\ref{eq:emu31}), depends 
on the combination $s^2_{14} s^2_{24}$. This term (symmetric under 
$\theta_{14} \leftrightarrow \theta_{24}$) dominates over the sub-leading
$\Delta m^2_{atm}$-dependent (non-symmetric) corrections. Therefore, 
the energy dependence of $N_{i,p}$ is not enough to resolve the two angles. 

\begin{figure}
\begin{center}
\begin{tabular}{cc}
\epsfxsize6.5cm\epsffile{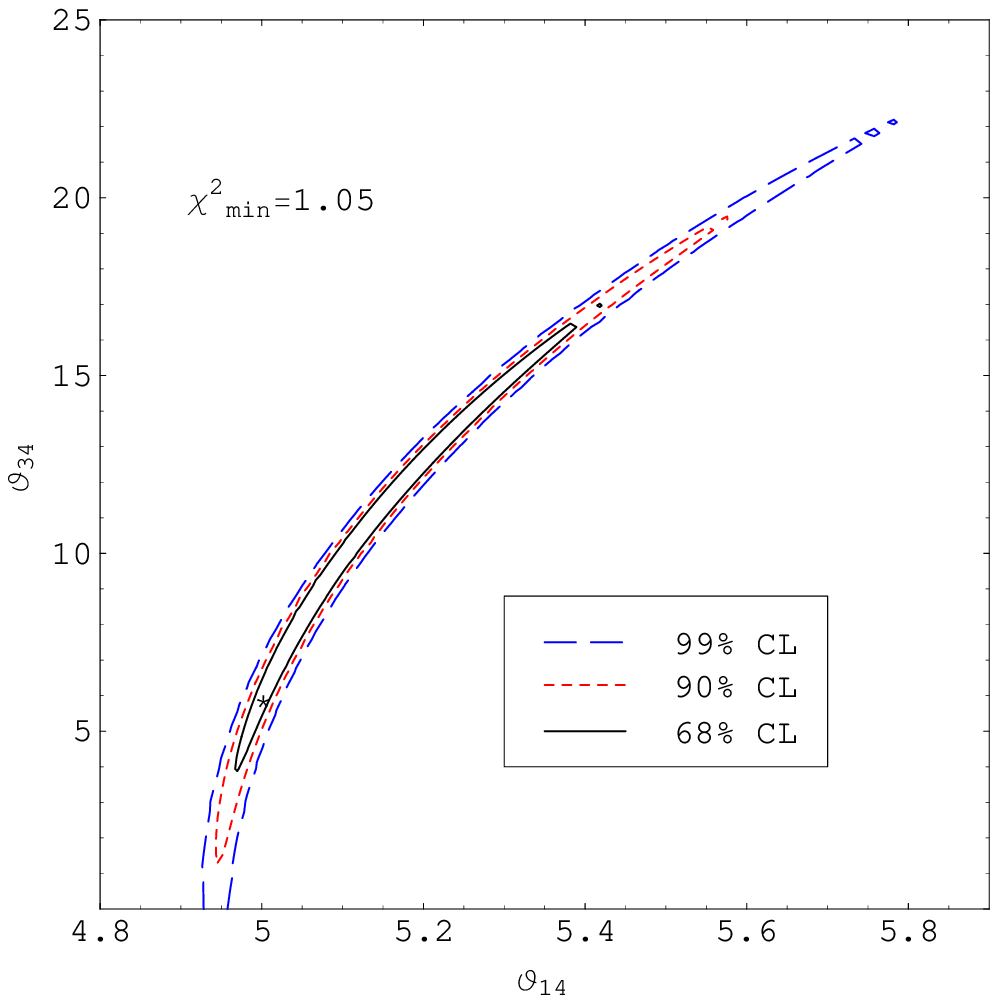} & 
\epsfxsize6.5cm\epsffile{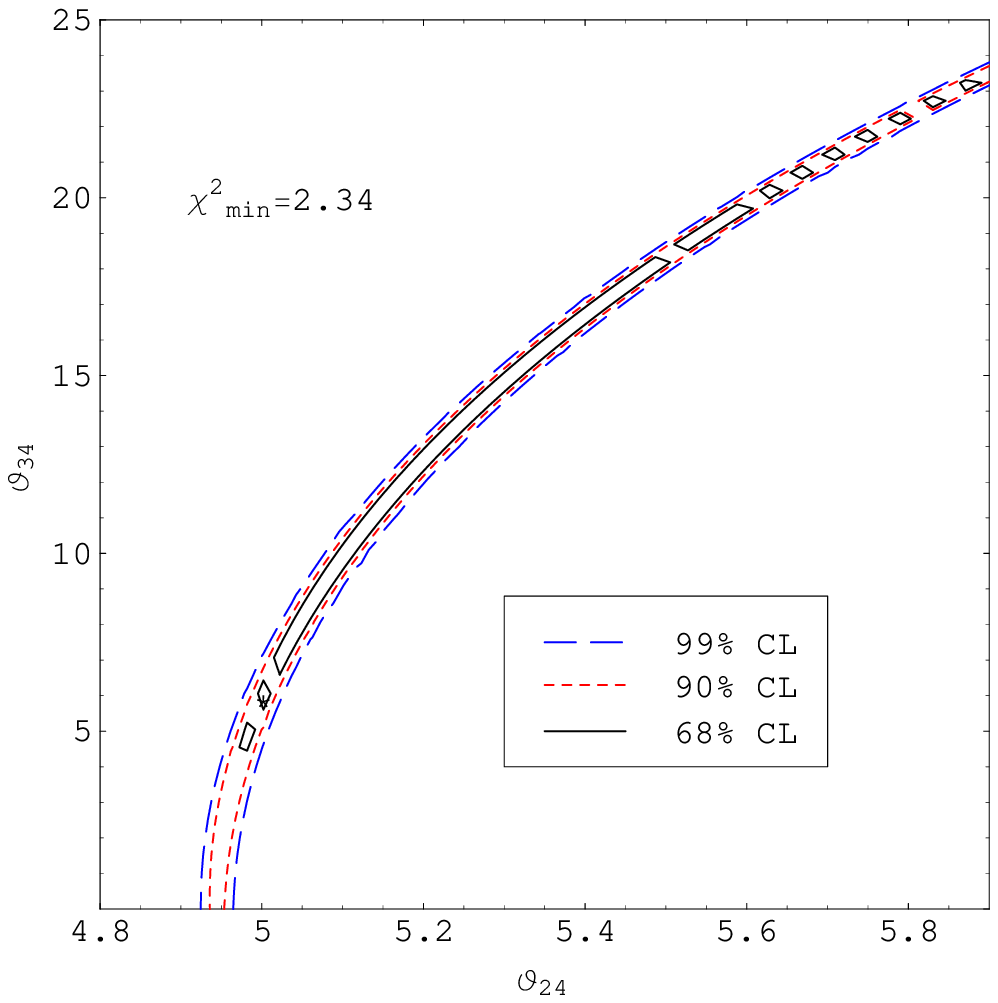}  \\
\end{tabular}
\caption{\it{68, 90 and 99 \% CL contours resulting from a $\chi^2$ fit of
$\theta_{14}$ and $\theta_{34}$ (left) or $\theta_{24}$ and $\theta_{34}$ (right). 
The parameters used to generate the ``data'' are depicted by a star:
$\theta_{14} = 5^\circ, \theta_{34}=6^\circ$ (left);
$\theta_{24} = 5^\circ, \theta_{34}=6^\circ$ (right). 
Only statistical errors are included.}} 
\label{fig:duetheta}
\end{center}
\end{figure}

Next, we consider the simultaneous measurement
of $\theta_{34}$ and one of $\theta_{14}, \theta_{24}$.
We compute the leading corrections to eq. (\ref{eq:emu31}) in powers of 
$\Delta m^2_{atm}$. 
For vanishing phases and 
$s_{13} = s_{24} = \epsilon \, , \, \theta_{23} = 45^\circ$, 
the leading terms are 
\bea
\PCPCTRE (\nu_e \to \nu_\mu) &=& 4 \epsilon^2 s_{14}^2 c_{34}^4 
               \,     \sin^2 \Delta_{34} 
- 2 c_{34}^2 s_{34} s_{14}^2 \ \epsilon \ \Delta_{23} \ \sin (2 \Delta_{34}) \nn \\
&+& O(\epsilon^4) + O (\Delta_{23}^2) + O ( \epsilon^2 \Delta_{23}) \, .
\label{eq:fig19left}
\eea
For vanishing phases and 
$s_{13} = s_{14} = \epsilon \, , \, \theta_{23} = 45^\circ$, 
the leading terms are, instead,
\be
\PCPCTRE (\nu_e \to \nu_\mu) = 4 \epsilon^2 c_{24}^2 s_{24}^2 c_{34}^4 
                  \,  \sin^2 \Delta_{34}
+ O(\epsilon^4) + O (\Delta_{23}^2) + O ( \epsilon^2 \Delta_{23}) \, .
\label{eq:fig19right}
\ee
In Fig.~\ref{fig:duetheta} (left) we present the 68, 90 and 99 \% confidence level 
contours for a simultaneous fit of $\theta_{14}$ and $\theta_{34}$ for a ``data'' set 
generated with $\theta_{14}=5^\circ, \theta_{24} = 2^\circ$ and $\theta_{34}=6^\circ$. 
This figure corresponds to eq. (\ref{eq:fig19left}), where the leading
correction to the one-mass dominance formula is $O(\epsilon \Delta_{23})$.
In Fig.~\ref{fig:duetheta} (right) we present the 68, 90 and 99 \% confidence level 
contours for a simultaneous fit of $\theta_{24}$ and $\theta_{34}$ for a ``data'' set 
generated with $\theta_{24}=5^\circ, \theta_{14} = 2^\circ$ and $\theta_{34}=6^\circ$. 
This figure corresponds to eq. (\ref{eq:fig19right}): notice that the 
$O(\epsilon \Delta_{23})$ correction to the one-mass dominance formula is absent 
and the leading corrections start at higher orders. The sensitivity to $\theta_{34}$ 
is therefore suppressed with respect to eq. (\ref{eq:fig19left}). 
In summary, $\theta_{14}$ or $\theta_{24}$ are reconstructed with
a precision of tenths of degree; on the contrary, $\theta_{34}$ is measurable with 
a very poor precision in both cases, 
with slightly better results in Fig. \ref{fig:duetheta} (left). 

\begin{figure}[h!]
\begin{center}
\epsfxsize7cm\epsffile{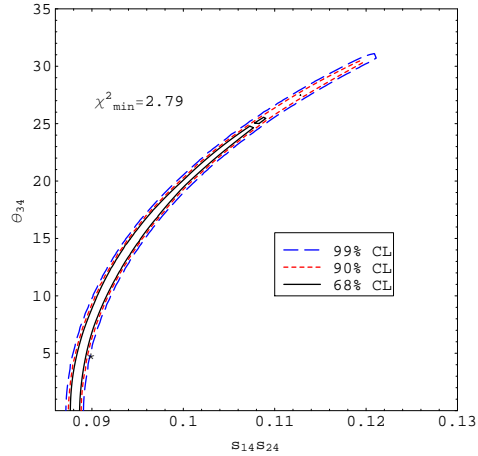} 
\caption{\it{68, 90 and 99 \% CL contours resulting from a $\chi^2$ fit of
$s_{14} s_{24}$ and $\theta_{34}$. The parameters used to generate the ``data'' are 
depicted by a star, $s_{14} s_{24} = 0.09, \theta_{34} = 5^\circ$. 
Only statistical errors are included.}} 
\label{fig:tretheta}
\end{center}
\end{figure}

We try also to simultaneously fit the combination $s_{14} s_{24}$ and $\theta_{34}$.
In Fig. \ref{fig:tretheta} we generate the ``data'' for 
$s_{14} s_{24} = 0.09, \theta_{34} = 5^\circ$. 
As in Fig.~\ref{fig:duetheta}, we observe that $\theta_{34}$ is poorly reconstructed,
whereas $s_{14}s_{24}$ is severely constrained.  
Our results seem to indicate that is very difficult to measure $\theta_{34}$
using the energy dependence of sub-leading effects in the $\nu_e \to \nu_\mu$ channel. 

\begin{figure}[h!]
\begin{center}
\epsfxsize7cm\epsffile{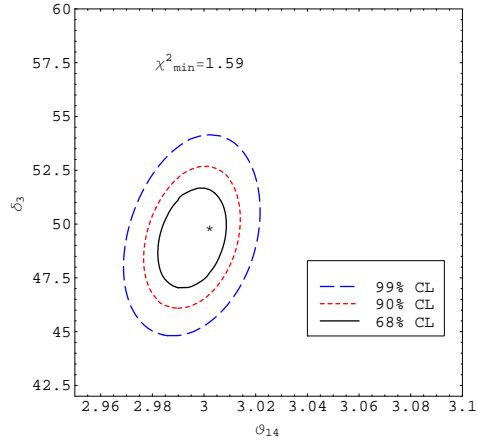} 
\caption{\it{68, 90 and 99 \% CL contours resulting from a $\chi^2$ fit of
$\theta_{14}$ and $\delta_3$ (left). The parameters used to generate the ``data'' 
are depicted by a star, $\theta_{14} = 3^\circ, \delta_3 = 50^\circ$. 
Only statistical errors are included.}} 
\label{fig:thetadelta}
\end{center}
\end{figure}

Finally, we consider the simultaneous measurement of one gap-crossing angles
(the other two being fixed at some small value) and one CP-violating phase. 
In this case, the leading correction to the one-mass dominance formula, 
eq. (\ref{eq:emu31}), is the CP-odd contribution, eq. (\ref{eq:emu31cp}).
In Fig.~\ref{fig:thetadelta}  we show the confidence level contours for 
a simultaneous fit of $\theta_{14}$ and $\delta_3$. The theoretical values
for which the ``experimental data'' have been generated are: 
$\theta_{14} = 3^\circ, \delta_3 = 50^\circ$ ($\theta_{24} = \theta_{34} = 2^\circ$). 
We see that the angle is reconstructed
with a precision of tenths of degree, whereas the phase is measured with
a precision of tens of degrees only. This is precisely the same situation 
of the three-family results of \cite{Cervera:2000kp}, however.

\newpage

\section{Conclusions}
\label{concl}

The ensemble of solar, atmospheric and LSND neutrino data can be explained with
three active plus one sterile flavour states. Although the four-family neutrino
mass spectrum preferred by the experimental data is the so-called 2+2 scheme
(with two almost degenerate pairs well separated by the LSND mass difference), 
the latest LSND result are marginally (at 99 \% CL) compatible with the 
3+1 scheme (three almost degenerate neutrinos separated from a fourth, mainly sterile, 
one). 

In this paper we studied the physical potential of a Neutrino Factory
in both four-family model schemes, in the spirit of \cite{Donini:1999jc,Donini:2000he}
where the 2+2 scheme was carefully examined. A re-analysis of the 2+2 results 
and a totally novel analysis in the 3+1 scheme has been presented. 
We have derived one- and two-mass scale dominance approximations appropriate 
for CP-even and CP-odd observables, respectively, in both schemes. 

The rich flavour content of a muon-decay based beam is extremely useful to 
determine or severely constrain the four-family model parameter space: 
in both schemes, the sensitivity to gap-crossing angles as small 
as $\sin^2 \theta_{ij} = 10^{-6}-10^{-4}$ (depending on the specific angle) 
can be achieved, with a 1 ton detector at $L \sim 1$ Km down from the source, 
for $n_\mu = 2 \times 10^{20}$ useful muons per year and 5 years of data taking. 
In the 3+1 scheme, we notice a slight loss in sensitivity with respect to the 2+2 scheme, 
that we interpret as a consequence of the higher power in the small gap-crossing angles
in the leading terms of the CP-conserving transition probabilities. 

CP violation may be easily at reach with a 10 Kton detector at $L = O (10)$ Km, 
especially through ``$\tau$ appearance'' signals, in both schemes. 
The increased significance of the CP-violating observables with respect
to the three-family model asymmetries is due to the fact that asymmetries proportional 
to $\Delta m^2_{atm}$ are possible, whereas in the three-family model the 
CP-violating asymmetries are proportional to the small $\Delta m^2_{sol}$. 
Moreover, in the four-family case the asymmetries are modulated by $\Delta m^2_{LSND}$, 
thus peaking at $L = O(10)$ Km; in three families they are modulated by 
$\Delta m^2_{atm}$ and therefore they peak at $L = O(1000)$ Km, 
thus significantly suffering from matter effects. 
On the contrary, matter effects are totally negligible in the 
four-family case. We give a simple argument based again on the power counting
of the small gap-crossing mixing angle to explain why in four-family models (both in the 
2+2 and 3+1 schemes) the $\nu_\mu \to \nu_\tau$ channel is the optimal one to study
CP-violating observables, whereas in three families $\nu_e \to \nu_\mu$ seems to be
best suited. 

Eventually, we consider a 10 Kton detector at $L = 40$ Km with only $\mu$ charge 
identification (of the magnetized iron type). In this case, we use the 
energy dependence of the wrong-sign muons to try to simultaneously reconstruct two
angles (or combinations of them) or a CP-violating phase and one
angle at a time. Our results suggest that it is not possible to constrain 
the whole CP-conserving parameter space using the $\nu_e \to \nu_\mu$ channel 
only. However, we can simultaneously measure one angle and one phase, 
with a precision of tens of degrees in the latter. 

Summarizing, a Neutrino Factory has an enormous 
discovery potential when four-neutrino models are considered, in both the 2+2 and 
3+1 schemes. If the LSND results will be confirmed by MiniBooNE, a muon storage 
ring appears to be an extremely powerful facility to perform precision measurement
on the whole four-family model parameter space and most probably the best opportunity
to discover CP violation in the leptonic sector. A (not-so-large) detector
with $\tau$-tracking and ($\mu, \tau$) charge identification capability is needed. 
If MiniBooNE will not confirm LSND, however, the 3+1 scheme will still represent 
a possible extension of the Standard Model and a Neutrino Factory can 
severely constrain its parameter space. This cannot be said of the 2+2 scheme, 
that would be ruled out by a negative result of MiniBooNE. 

\section*{Acknowledgements}

We acknowledge useful conversations with M.B. Gavela, J.J. Gomez Cadenas,
P. Hernandez, P. Lipari and S. Rigolin. We are particularly indebted with 
M. Lusignoli for discussions on many different aspects of this paper.

\end{document}